\documentclass[final,3p,times]{elsarticle}

\usepackage{amssymb}
\usepackage{amsmath}
\usepackage{physics}
\usepackage[bb=dsserif]{mathalpha}
\usepackage{bm}
\usepackage[colorlinks,allcolors=blue]{hyperref}
\usepackage[compat=1.1.0]{tikz-feynman}

\DeclareMathOperator{\openone}{\mathbb{1}}

\journal{Annals of Physics}

\begin{document}

\begin{frontmatter}



\title{Observable Measurement-Induced Transitions}


\author[label1,label2,label3]{Aleksei Khindanov\corref{cor1}} 
\ead{akhin@ameslab.gov}
\author[label1]{Igor L. Aleiner}
\author[label1]{Lara Faoro}
\author[label1]{Lev B. Ioffe}
\cortext[cor1]{Corresponding author}

\affiliation[label1]{organization={Google Quantum AI},
            city={Santa Barbara},
            state={CA},
            country={USA}}
\affiliation[label2]{organization={Department of Physics, University of California, Santa Barbara},
            state={CA},
            postcode={93106},
            country={USA}}
\affiliation[label3]{organization={Ames National Laboratory, U.S. Department of Energy},
            city={Ames},
            state={IA},
            postcode={50011},
            country={USA}}

\begin{abstract}
One of the main postulates of quantum mechanics is that measurements destroy quantum coherence (wave function collapse). Recently it was discovered that in a many-body system dilute local measurements still preserve some coherence across the entire system. As the measurement density is increased, a phase transition occurs that is characterized by the disentanglement of different parts of the system. Unfortunately, this transition is impossible to observe experimentally for macroscopic systems because it requires an exponentially costly full tomography of the many-body wave function or a comparison with the simulation on an oracle classical computer. In this work we report the discovery of another measurement-induced phase transition that can be observed experimentally if quantum dynamics can be reversed. On one side of this phase transition the quantum information encoded in some part of the Hilbert space is fully recovered after the time inversion.  On the other side, all quantum information is corrupted. This transition also manifests itself as the change in the behavior of the probability to observe the same measurement outcome in the process that consists of identical blocks repeated many times. In each block the unitary evolution is followed by the measurement. On one side of the transition the probability decreases exponentially with the number of repetitions, on the other it tends to a constant as the number of repetitions is increased. We confirm the existence of the proposed phase transition through numerical simulations of realistic quantum circuits and analytical calculations using an effective random-matrix theory model.
\end{abstract}




\end{frontmatter}




\section{\label{sec:Intro}Introduction}

Recently, there has been a considerable interest in problems involving a many-body system under unitary evolution with local measurements.
A celebrated example is a class of transitions known as the measurement-induced phase transitions (MIPTs)~\cite{Potter2022, Fisher2023}.
Unfortunately, even though those transitions are readily observed numerically (i.e. they are \textit{computable}), they are almost impossible to detect in an experimental many-body system or on a large-scale quantum computer.
The goal of this paper is to propose a protocol for a quantum computer to make MIPTs \textit{observable}.
Before we proceed, let us formulate the difference between observable and computable quantities.

An observable in quantum mechanics is a physical quantity represented by a local Hermitian operator $O$: the eigenvalues $o_{\alpha}$ of $O$ correspond to the possible outcomes of measuring the observable, and the eigenvectors represent the states in which the observable has a definite value.
Due to locality of the operator $O$, in a many-body system there are still exponentially many states corresponding to a single eigenvalue $o_{\alpha}$ -- these states $|\psi_{\alpha,i}\rangle$ form a complete basis in a Hilbert subspace describing the effect of the measurement.
For the case of projective measurements the density matrix transforms under the measurement as $\rho \to \rho_{\alpha} = P_{\alpha}\rho P_{\alpha}$, where $P_{\alpha}=\sum_i |\psi_{\alpha,i}\rangle \langle \psi_{\alpha,i}|$ is a projection operator, while the probability of a measurement outcome $o_{\alpha}$ is given by $w_{\alpha}=\Tr[\rho_{\alpha}]$.
Despite the measurement, the density matrix $\rho_{\alpha}$ still spans an exponentially large Hilbert subspace, and it is possible to study a subsequent unitary evolution of the system while using $\rho_{\alpha}$ as the initial condition, resulting in a density matrix $\rho_{\alpha}^U = U\rho_{\alpha} U^{\dag}$, where $U$ is the evolution operator.
The system then can be measured again, yielding measurement outcome $o_{\beta}$ and the density matrix $\rho_{\alpha,\beta} = P_{\beta}\rho_{\alpha}^U P_{\beta}$.
This process of performing combined unitary evolution and measurements can be continued, generating the density matrix $\rho_{\alpha,\beta,...}$ corresponding to a particular quantum trajectory $\{\alpha,\beta,...\}$.
The probability of the trajectory is given by $w_{\alpha,\beta,...}=\Tr[\rho_{\alpha,\beta,...}]$.
It is this probability that any quantum mechanical theory can predict and its predictive power can be checked by a statistical analysis of experimental outcomes.
It is important to emphasize that this probability is a quantity linear in the density matrix.

\begin{figure}[t]
    \centering
    \includegraphics[height = 0.15\columnwidth]{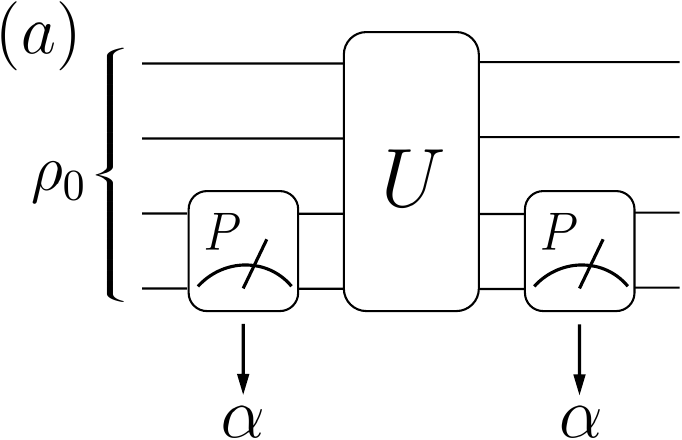}\qquad\quad
    \includegraphics[height = 0.15\columnwidth]{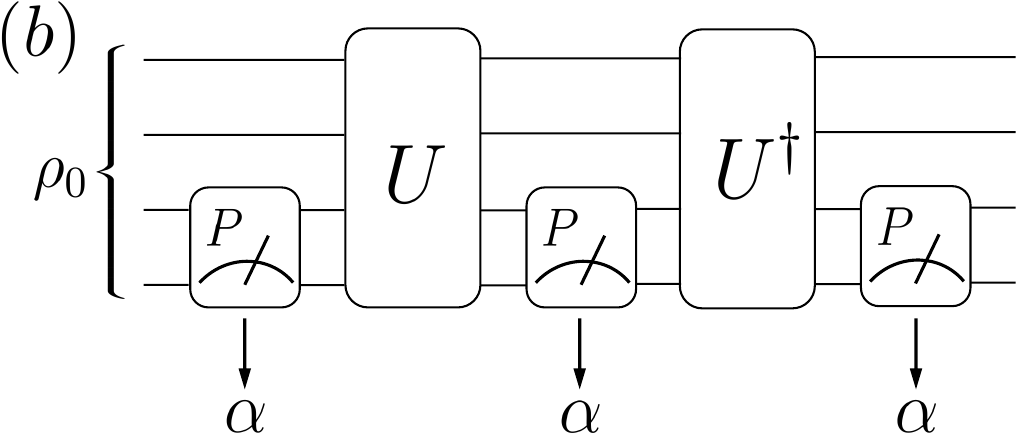}\\
    \vspace{0.1cm}
    \includegraphics[height = 0.15\columnwidth]{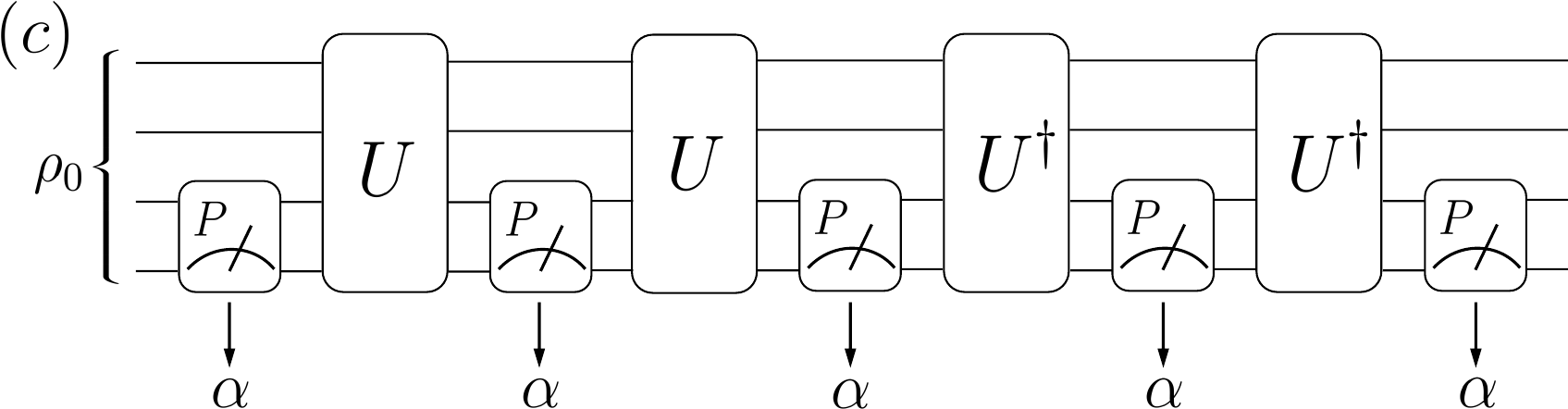}
	\caption{Schematic depiction of quantum circuits involving measurements (represented by a projector $P$) and unitary evolution $U$. Circuits (b) and (c) also involve a reversed unitary evolution $U^{\dag}$. Only specific measurement outcome (which we denote $\alpha$) is accepted throughout the measurements.}
    \label{fig:circuit}
\end{figure}

This linearity has to be contrasted to computable quantities usually studied in the context of MIPTs, which are written in terms of a normalized density matrix:
\begin{equation}
    \varrho _{\alpha,\beta,...} = \frac{\rho_{\alpha,\beta,...}}{w_{\alpha,\beta,...}}.
\end{equation}
Those computables are in general averaged over all trajectories:
\begin{equation}
    \bar{\mathcal F} = \sum_{\alpha,\beta,...}w_{\alpha,\beta,...}  \mathcal F(\varrho _{\alpha,\beta,...}).
    \label{eq:F_bar}
\end{equation}
If function $\mathcal F$ is linear, then $\bar{\mathcal F}$ is an observable quantity.
However, all quantities which characterize MIPTs are nonlinear by construction.
These include purity~\cite{Gullans2020_1,Fidkowski2021}, where $\mathcal F (\varrho)=\Tr[\varrho^2]$, von Neumann entropy, where $\mathcal F (\varrho)=-\Tr[\varrho \log\varrho]$, or the entanglement entropy~\cite{Li2018,Skinner2019,Li2019}, where the system is divided into two subregions, $A$ and $\bar A$, and the entropy is written in terms of a reduced density matrix: $\mathcal F (\varrho)=-\Tr[\varrho_A\log\varrho_A ]$, where $\varrho_A=\Tr_{\bar A}\varrho$.
Measuring all these quantities even for a single trajectory requires performing a complete tomography of the full density matrix and calculating quantities on classical machines, which demands exponential resources.
Moreover, the summation over trajectories in Eq.~\eqref{eq:F_bar} brings in additional exponential overhead.
It was recently suggested~\cite{Li2023,Hoke2023,Garratt2024,Kamakari2024} that MIPTs can be observed by studying cross-correlations between experimental results and data generated by a classical computer, similarly to the Google's quantum supremacy experiment~\cite{Arute2019}.
However, this method relies on classical simulations, and thus it is either not scalable to large system sizes (at least in the phase characterized by the volume-law entanglement), or have to address a special class of simulatable quantum circuits (such as Clifford circuits).

\begin{figure}[t]
    \centering
    \includegraphics[width = 0.31\columnwidth]{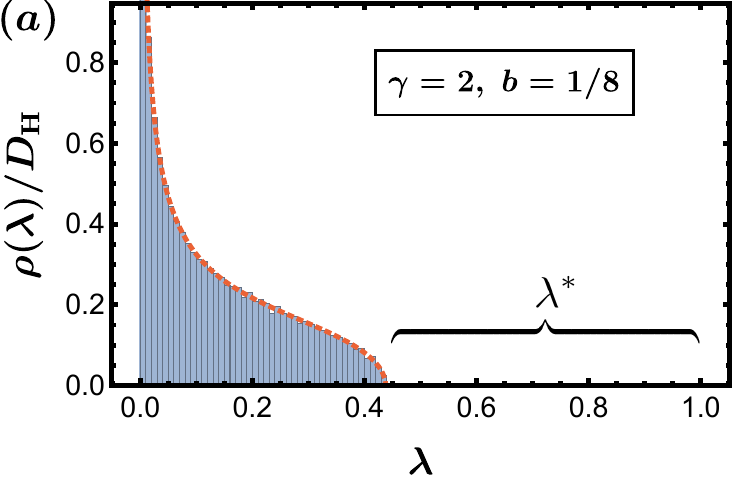}\quad
    \includegraphics[width = 0.31\columnwidth]{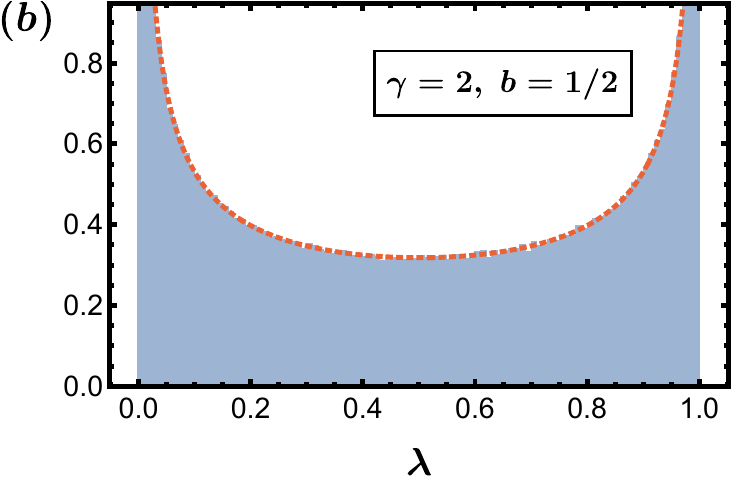}\quad
    \includegraphics[width = 0.31\columnwidth]{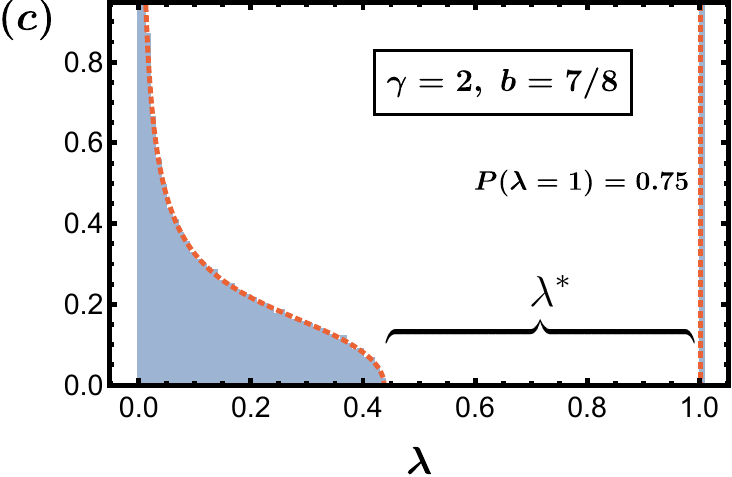}
    \caption{Histograms representing spectral density of a numerically generated random matrix $\Lambda_P$ of Eq.~\eqref{eq:Lambda_P} with the random unitary $U$ drawn from the ensemble \eqref{eq:GUE}-\eqref{eq:cayley} with $\gamma=2$. The dimension of the Hilbert space is $D_{\rm H}=2048$. Averaging is performed over 40 random matrix realizations. For $b>1/2$ there is a finite density of eigenvalues at $\lambda=1$ given by $P(\lambda=1)$ defined in Eq.~\eqref{eq:P_lambda_1}. The red curve shows the analytical expression \eqref{eq:rho_lambda_math} for the spectral density of $\Lambda_P$ derived in Ref.~\cite{Collins2005} for the case when $U$ is Haar-random. Thus, the results for the ensemble \eqref{eq:GUE}-\eqref{eq:cayley} with $\gamma=2$ coincide with the results for the Haar-random ensemble.
	}
    \label{fig:W_eval_hist}
\end{figure}

In this work we suggest an alternative MIPT, the observability of which does not rely on classical simulations and thus is scalable.
The transition is evident in the properties of an operator 
\begin{equation}
    \Lambda_P = P U^{\dagger} P U P
    \label{eq:Lambda_P}
\end{equation}
[see Fig.~\ref{fig:circuit}(b)], where $U$ is a unitary evolution operator and $P$ is a projector ($P^2=P$) corresponding to a fixed measurement outcome. Superficially similar operators were studied in Ref.~\cite{Turkeshi2024}, where instead of the middle projector the authors used a quantum channel characterizing either coherent or incoherent errors. The phase transitions analyzed in Ref.~\cite{Turkeshi2024}, however, are conceptually different from the ones studied in this paper.
We characterize the measurement by the rank of the projector: $\Tr[P]=bD_{\rm H}$, where $D_{\rm H}$ is the dimensionality of the Hilbert space [$\log(D_{\rm H})$ is an extensive quantity], while $b$ is a non-extensive parameter.
For example, measuring one qubit and accepting only one outcome leads to $b=1/2$.
For a two-qubit measurement (where there are four total possible outcomes), accepting only one measurement outcome corresponds to $b=1/4$, whereas accepting any of the three measurement outcomes without distinguishing them yields $b=3/4$. Such acceptance can be achieved in practice by performing gates with an ancilla qubit, for instance by applying the CCNOT gate to the two data qubits (as control qubits) and the ancilla (as the target qubit), measuring the ancilla and accepting only outcome ``1".
In general, $b$ can be made almost continuous by increasing the number of measured qubits.

The operator $\Lambda_P$ is Hermitian and its eigenvalues $\lambda_i$ are confined between $\lambda=0$ and $\lambda=1$.
For $b=0$, all eigenvalues of $\Lambda_P$ are equal to zero, whereas for $b=1$ all its eigenvalues are equal to one.
We have investigated the spectrum of $\Lambda_P$ analytically for a random unitary $U$ and studied it numerically for $U$ obtained by the repeated application of the gates on a realistic quantum processor that mimic random unitary.
We discovered that in all cases that we studied the finite density of eigenvalues at $\lambda=1$ persists for $b<1$ as long as $b>b_c$, where $b_c=1/2$ for the analytically solvable model and $b_c\approx 1/2$ for the evolution on one- and two-dimensional grids. The eigenvalues at $\lambda=1$ are separated from the rest by a finite gap at $b\neq b_c$  for the analytically solvable model and the 1D chain, while numerically studied evolution on the 2D lattice shows suppression of the spectral density. 
At $b=b_c$ the gap closes, and for $b<b_c$ all eigenvalues of $\Lambda_P$ are strictly less than $\lambda=1$. We conjecture that the difference between the analytically solvable random-matrix model and the evolution on the 2D grid stems from the finite size effects of the latter. 
The described transition in the spectrum is depicted in Fig.~\ref{fig:W_eval_hist} for a numerically generated random matrix $U$.

On the first glance, this transition looks like a mathematical curiosity and cannot be accessed experimentally.  
Indeed, the simplest way to construct $\Lambda_P$ would be to start with some initial density matrix $\rho_0$, perform a measurement accepting a certain outcome (we call it $\alpha$), perform unitary evolution and then perform a second measurement accepting the same outcome $\alpha$ [see Fig.~\ref{fig:circuit}(a) for a schematic depiction of the procedure]. This process would yield a quantum trajectory whose probability is given by $w_{\alpha\alpha}=\Tr[\Lambda_P\rho_0]$. However, this probability would not reveal the transition shown in Fig.~\ref{fig:W_eval_hist} because the trace operation involves a summation over all eigenvalues so that the feature at $\lambda=1$ is integrated out.

Nevertheless, recent advances in reversing the direction of time dynamics on a quantum computer \cite{Mi2021} have made the observation of the phase transition possible. Note that the ability to invert the sign of the dipole-dipole interaction in the secular approximation is a well established NMR technique \cite{Rhim1971} and can be possibly applied to this problem as well.
The ability to reverse the direction of time dynamics allows us to study the following process: starting with some initial density matrix $\rho_0$ we perform a measurement while accepting an outcome $\alpha$, perform unitary evolution, then perform a second measurement accepting the same outcome  $\alpha$, run the same unitary evolution backwards, and afterwards conduct another measurement while accepting the outcome  $\alpha$.
This process is schematically depicted in Fig.~\ref{fig:circuit}(b).
The probability of the resulting quantum trajectory in this case is given by $w_{\alpha\alpha\alpha}=\Tr[PU^{\dag}PUP\rho_0PU^{\dag}PUP]\equiv\Tr[\Lambda_P^2\rho_0]$.
We can further continue this procedure running the evolution back and forth, generating the trajectory with a probability given by
\begin{equation}
    w(d)=w_{\underbrace{\scriptstyle \alpha\alpha\dots\alpha }_\text{$(d+1)$}}=\Tr[\Lambda_P^{d}\rho_0],
\end{equation}
where $d$ is the number of times the unitary $U$ and its time-reversal counterpart $U^{\dag}$ are applied throughout the process; the number of corresponding Floquet cycles is Int[$d/2$].

Note that $w(d)$ is proportional to the Laplace transform of the density of states in Fig.~\ref{fig:W_eval_hist}, and its asymptotic behavior in the limit $d\gg1$ is given by
\begin{subequations}\label{eq:w_d}
\begin{align}
    w(d) &\sim \exp[-\lambda^{\ast}(\delta_b)d],\quad \text{   for ${\delta_b<0}$}, \label{eq:w_d_1}  \\
    w(d) &\sim \frac{1}{d^{\eta}},\quad \text{for ${\delta_b=0}$}, \label{eq:w_d_0}  \\
    w(d) &\sim w_{\infty}(\delta_b) + \exp[-\lambda^{\ast}(\delta_b)d],\quad \text{for ${\delta_b>0}$} ,\label{eq:w_d_2}
\end{align}
\end{subequations}
where $\delta_b = b-b_c$.
The first term in Eq.~\eqref{eq:w_d_2} is proportional to the number of the eigenvalues at $\lambda=1$, while $\lambda^{\ast}(\delta_b)$ corresponds to the gap in the spectrum (see Fig.~\ref{fig:W_eval_hist}).
Behavior \eqref{eq:w_d} is a hallmark of a phase transition as an analytic crossover between asymptotic behaviors of the ``correlation functions" \eqref{eq:w_d_1},\eqref{eq:w_d_2} is not possible.
More details about functions $\lambda^{\ast}(\delta_b)$, $w_{\infty}(\delta_b)$ as well as the critical exponent $\eta$ are presented in Sections~\ref{sec:num_analyt} and \ref{sec:RMT}.
Here we also would like to point out that an exponential decay of the probability similar to \eqref{eq:w_d_1} has been recently derived in other random-matrix models of monitored dynamics~\cite{Bulchandani2024}, while a polynomial postselection overhead in MIPTs have been recently investigated~\cite{Passarelli2024,Delmonte2024} for various infinite-range spin models.
We emphasize, however, that the models and the transitions studied in Refs.~\cite{Bulchandani2024,Passarelli2024,Delmonte2024} are distinct from the ones presented in this paper.

The remainder of this paper is organized as follows.
Section~\ref{sec:num_analyt} gives the qualitative arguments and numerical results for both random unitary and realistic evolutions that support the existence of the phase transition.
Section~\ref{sec:RMT} presents an analytically tractable random-matrix model describing the details of the phase transition.
Section \ref{sec:Conclusion} contains concluding remarks.

\section{\label{sec:num_analyt}Qualitative arguments for the existence of the transition and numerical results}

\subsection{\label{sec:qualit_arg}Qualitative arguments}

To analyze $\Lambda_P$, we notice that the operator ${P_U = U^{\dag}PU}$ is also a projection operator ($P_U = P_U^2$) with the same rank as $P$: $\Tr[P_U]=bD_{\rm H}$, $\Lambda_P = PP_UP$.
For Haar-random unitaries $U$, Ref.~\cite{Collins2005} studied the operator $\Lambda_P$ as a product of random projectors using free probability theory, revealing the transition in its spectrum depicted in Fig.~\ref{fig:W_eval_hist} (see Section~\ref{sec:numerics} for a more detailed description of the results obtained in Ref.~\cite{Collins2005}).
Here we present more general, qualitative arguments for the existence of the transition.

For convenience, we consider a basis in which $P$ is diagonal.
In this basis $P$ has a block-diagonal form 
\begin{equation}
	P =
	\begin{pmatrix}
		\openone_{bD_{\rm H}\times bD_{\rm H}} & 0 \\
		0 & 0_{\bar{b}D_{\rm H}\times \bar{b}D_{\rm H}}
	\end{pmatrix}
    \label{eq:P_mx}
\end{equation}
where $\bar{b}\equiv 1-b$, while $P_U$ is not block-diagonal. 
Rewriting $P_U$ as
\begin{equation}
    P_U = \openone_{D_{\rm H}\times D_{\rm H}} - Q_{D_{\rm H}\times D_{\rm H}},
    \label{eq:P_U}
\end{equation}
we note that $Q$ is also a projection operator with $\text{rank}[Q] = \bar{b}D_{\rm H}$.
It can be further decomposed as
\begin{equation}
    Q =\sum_{i=1}^{\bar{b}D_{\rm H}} |\phi_{i}\rangle \langle \phi_{i}|,
\end{equation}
where $ \langle \phi_{i}|\phi_{j}\rangle = \delta_{ij}$.
Substituting the above representations of $P$ and $P_U$ into the expression for $\Lambda_P$, we find that $\Lambda_P$ also has a block-diagonal form:
\begin{equation}
	\Lambda_P =
	\begin{pmatrix}
		[\openone - \tilde{Q}]_{bD_{\rm H}\times bD_{\rm H}} & 0 \\
		0 & 0_{\bar{b}D_{\rm H}\times \bar{b}D_{\rm H}}
	\end{pmatrix}
\end{equation}
where
\begin{equation}
    \tilde{Q} =\sum_{i=1}^{\bar{b}D_{\rm H}} |\tilde\phi_{i}\rangle \langle \tilde\phi_{i}|, \quad |\tilde\phi_{i}\rangle= P|\phi_{i}\rangle.
    \label{eq:Q_tilde}
\end{equation}
However, unlike $\{|\phi_{i}\rangle\}$, $\{|\tilde\phi_{i}\rangle\}$ are not orthogonal to each other and are not normalized.
Let us assume that for $\bar{b}<b$ $\{|\tilde\phi_{i}\rangle\}$ remain linearly independent (due to the absence of the correlation between $P$ and $U$).
Then, $\text{rank}[\tilde{Q}]=\bar{b}D_{\rm H}$ and all its nonzero eigenvalues $\tilde q_j$ are confined, $0<\tilde q_j<1$ for $(b-\bar{b})D_{\rm H}< j\leq bD_{\rm H}$.  In the basis of  $\tilde{Q}$ eigenvectors  $\Lambda_P$ is given by
\begin{equation}
	\Lambda_P =
	\begin{pmatrix}
		\openone_{(b-\bar{b})D_{\rm H}\times (b-\bar{b})D_{\rm H}} & 0 & 0 \\
		0 & \Upsilon_{\bar{b}D_{\rm H}\times \bar{b}D_{\rm H}} & 0 \\
        0 & 0 & 0_{\bar{b}D_{\rm H}\times \bar{b}D_{\rm H}}
	\end{pmatrix},
    \label{eq:Lambda_P_mx}
\end{equation}
where $\Upsilon = \text{diag}[1-\tilde q_j]$.
This  argument shows that for $b>\bar b$, or, equivalently, $b>b_c=1/2$, the operator $\Lambda_P$ has $(2b-1)D_{\rm H}$ eigenvalues at $\lambda=1$.
Hence, in Eq.~\eqref{eq:w_d_2} $w_{\infty}(\delta_b)\propto \delta_b$ for $\delta_b>0$.

We emphasize that this argument works only if $b>b_c$. For $b<b_c$, the number of eigenfunctions $\{|\tilde\phi_{i}\rangle\}$ in the sum \eqref{eq:Q_tilde} becomes larger than the dimensionality $b D_{\rm H}$ of the Hilbert subspace spanned by the eigenfunctions of $P$, and they are no longer linearly independent.
The appropriate course of action in this case is to use, instead of Eq.~\eqref{eq:P_U}, a direct decomposition of the operator $P_U$:
\begin{equation}
    P_U =\sum_{i=1}^{bD_{\rm H}} |\varphi_{i}\rangle \langle \varphi_{i}|.
\end{equation}
Then, following the same steps as above, we obtain the operator $\Lambda_P$ in the following form:
\begin{equation}
	\Lambda_P =
	\begin{pmatrix}
		\Omega_{bD_{\rm H}\times bD_{\rm H}} & 0 \\
        0 & 0_{\bar{b}D_{\rm H}\times \bar{b}D_{\rm H}}
	\end{pmatrix},
\end{equation}
where $\Omega = \text{diag}[\omega_j]$ and $0<\omega_j<1$.
This shows that for $b<b_c$, the operator $\Lambda_P$ does not possess a finite density of eigenvalues at $\lambda=1$, and, together with the results for $b>b_c$, proves the existence of the transition at $b=b_c$.
However, further details of the spectrum of $\Lambda_P$, for example the size of the gap, cannot be determined using this simple argument.
To do this, we further employ analytical and numerical techniques that will be presented in the remainder of the paper.

\subsection{The phase transition does not imply the existence of a unitary subspace}

One can view the Hermitian operator, $\Lambda_P$,  as a product:
\begin{equation}
    \Lambda_P = W_P^{\dagger}W_P, \quad W_P = PUP,
\end{equation}
where the operator $W_P$ is a contraction because 
${\lVert W_P \rVert \leq 1}$. 
It might be tempting to interpret the finite density of $\lambda=1$ eigenvalues as a decomposition of the  Hilbert space into two independent subspaces, such that in one subspace, $\cal{D}$, the decay occurs, while in the other, $\cal{U}$, the dynamics remains unitary. Obviously, the reverse is true: if such decomposition takes place, there is a finite density of  $\lambda=1$ eigenvalues. Furthermore, if the operator $W_P$ is unitary in some subspace, the product $W_P^{\dagger n} W_P^n$ is the identity in this space for any $n$ and the defect operator defined by ${D_n=\left( \openone-W_P^{\dagger  n}W_P^n\right)^{1/2}}$ is zero in this space. It turns out \cite{Nagy} that for any contraction operator there exists a canonical decomposition of the Hilbert space and the unitary subspace is uniquely given by the intersection of the kernels of the defect operators: ${{\cal U}={\bigcap_{n=0}^{n=\infty} \text{Ker} D_n  \cup \text{Ker} D_n^{\dagger}}}$.

To determine if  $\cal{U}$  is non-empty in the thermodynamic limit, we introduce
\begin{equation}
    \Lambda_P^{(n)} = (W_P^{\dagger})^nW_P^n
\end{equation}
and analyze its spectrum. We find that the dimension of the space in which $\lambda=1$ for the operator  $\Lambda_P^{(n)}$ goes to zero as $n\rightarrow \infty$. As a result, the intersection of the kernels of the defect operators that gives $\cal{U}$ is certainly empty.  

We start by considering the case of $n=2$.
In this case,
\begin{equation}
    U\Lambda_P^{(2)}U^{\dag} = P_{U^{\dag}}\Lambda_PP_{U^{\dag}}, \quad P_{U^{\dag}}= UPU{^\dag}.
\end{equation}
Let us work in the basis where $P_{U^{\dag}}$ has a diagonal form \eqref{eq:P_mx}.
In this basis we rewrite $\Lambda_P$ as
\begin{align}
    \Lambda_P =\openone_{D_{\rm H}\times D_{\rm H}} - \sum_{i=1}^{2\bar b D_{\rm H}}  q_i^{(2)}|\phi^{(2)}_i\rangle\langle\phi^{(2)}_i|,
\end{align}
where $0<  q_i^{(2)}\leq1$.
Counting the ranks of the second term and employing the same assumption of the linear independence of the wavefunctions $|\tilde\phi^{(2)}_i\rangle = P_{U^{\dag}}|\phi^{(2)}_i\rangle$, we obtain, similarly to Eq.~\eqref{eq:Lambda_P_mx},
\begin{equation}
	\Lambda_P^{(2)} =
	\begin{pmatrix}
		\openone_{(b-2\bar{b})D_{\rm H}\times (b-2\bar{b})D_{\rm H}} & 0 & 0 \\
		0 & \Upsilon^{(2)}_{2\bar{b}D_{\rm H}\times 2\bar{b}D_{\rm H}} & 0 \\
        0 & 0 & 0_{\bar{b}D_{\rm H}\times \bar{b}D_{\rm H}}
	\end{pmatrix}.
\end{equation}
Here $\Upsilon^{(2)} = \text{diag}[ \lambda_j^{(2)}]$ and $0<\lambda_j^{(2)}<1$.
Hence, for $b>b_c^{(2)}=2/3$, the operator $\Lambda_P^{(2)}$ has $(3b-2)D_{\rm H}$ eigenvalues at $\lambda=1$.
The case of arbitrary $n$ can be analyzed in a similar fashion, yielding
\begin{equation}
	\Lambda_P^{(n)} =
	\begin{pmatrix}
		\openone_{(b-n\bar{b})D_{\rm H}\times (b-n\bar{b})D_{\rm H}} & 0 & 0 \\
		0 & \Upsilon^{(2)}_{n\bar{b}D_{\rm H}\times n\bar{b}D_{\rm H}} & 0 \\
        0 & 0 & 0_{\bar{b}D_{\rm H}\times \bar{b}D_{\rm H}}
	\end{pmatrix},
\end{equation}
and the transition occurs at 
\begin{equation}
    b_c^{(n)}=n/(n+1).
    \label{eq:b_c^n}
\end{equation}
It follows from Eq.~\eqref{eq:b_c^n} that a decomposition of the Hilbert space into a nonunitary and a unitary part in fact \textit{does not occur}.
Note that the transition in the spectrum of $\Lambda_P^{(n)}$ can also be in principle observed for arbitrary $n$, similarly to the transition for $n=1$. For instance, one cycle of the corresponding circuit for $n=2$ is depicted in Fig.~\ref{fig:circuit}(c). In this way one can check experimentally for the presence of a subspace in which quantum information is preserved for realistic unitaries for which analytical solution is not available.

\subsection{\label{sec:numerics}Numerical results}

\subsubsection{Random-matrix}

We start by presenting numerical results for the spectrum of $\Lambda_P$ when $U$ is a randomly-generated $D_{\rm H}\times D_{\rm H}$ unitary matrix.
These results are illustrated as histograms representing the spectral density of $\Lambda_P$ in Fig.~\ref{fig:W_eval_hist}.
Here $U$ is drawn from an ensemble described later by Eqs.~\eqref{eq:GUE}-\eqref{eq:cayley} with $\gamma=2$.
In this section we show that for this value of $\gamma$ the spectral density coincides with the one obtained for the Haar-random ensemble.
Numerical results for values of $\gamma\neq 2$ are presented later in Section~\ref{sec:RMT}.

Histograms in Fig.~\ref{fig:W_eval_hist} reveal that for $b<1/2$ [see panel (a)] the spectrum is continuous and it is separated from $\lambda=1$ by a finite gap $\lambda^{\ast}$.
This gap closes at $b=1/2$ [see Fig.~\ref{fig:W_eval_hist}(b)] when the continuum spans the entire segment between $\lambda=0$ and $\lambda=1$.
For $b>1/2$ [see Fig.~\ref{fig:W_eval_hist}(c)] the continuum is again separated from $\lambda=1$ by the gap $\lambda^{\ast}$, but in contrast to the case of $b<1/2$, here there is a finite density of eigenvalues at $\lambda=1$ present in the spectrum.
We denote this density as
\begin{equation}
    P(\lambda=1) = \lim_{\varepsilon\to0}\int_{1-\varepsilon}^{1+\varepsilon} d\lambda \frac{\rho(\lambda)}{D_{\rm H}},
    \label{eq:P_lambda_1}
\end{equation}
where $\rho(\lambda)$ is the spectral density; in the case depicted in Fig.~\ref{fig:W_eval_hist}(c) for $b=7/8$ we have $P(\lambda=1)=3/4$.
The probability $w_{\infty}$ of Eq.~\eqref{eq:w_d_2} can be expressed through $P(\lambda=1)$ as
\begin{equation}
    w_{\infty} = \frac{1}{b}\Tr[P\rho_0]P(\lambda=1),
    \label{eq:w_inf_P_1}
\end{equation}
where $\Tr[\rho_0]=1$.
In addition, we note that the continuum spectrum for ${b=1/2+\Delta b}$, $\Delta b>0$, and ${b=1/2-\Delta b}$ is identical.

The spectrum of a random matrix $\Lambda_P$ has been studied analytically in Ref.~\cite{Collins2005} for the case when $U$ is a Haar-random unitary.
Treating $\Lambda_P$ as a product of random projectors and employing free probability theory, Ref.~\cite{Collins2005} obtained the following expression for the spectral density:
\begin{align}
    \frac{\rho(\lambda)}{D_H}= \bar b\delta(\lambda) + (b-\bar b)\theta(b-\bar b)\delta(\lambda-1)+\frac{1}{2\pi(1-\lambda)}\sqrt{\frac{4b\bar b-\lambda}{\lambda}}\theta(\lambda)\theta\left[4b\bar b-\lambda\right],
    \label{eq:rho_lambda_math}
\end{align}
where $\bar b=1-b$, $\theta(x)$ is the Heaviside step function and $\delta(x)$ is the Dirac delta function.
From Eq.~\eqref{eq:rho_lambda_math} it follows that the gap in the spectrum and the density of eigenvalues at $\lambda=1$ are given by
\begin{equation}
    \lambda^{\ast}=4b\bar b,\quad P(\lambda=1)=b-\bar b,
\end{equation}
respectively.
Function \eqref{eq:rho_lambda_math} is plotted in Fig.~\ref{fig:W_eval_hist} alongside the numerically computed histograms, revealing the perfect agreement between our numerical results and the analytical conclusions of Ref.~\cite{Collins2005}.

Figure~\ref{fig:W_eval_hist} demonstrates that there is a transition present in the spectrum of $\Lambda_P$ at a critical point $b=b_c=1/2$.
This spectral transition gives rise to the behavior of the quantum trajectory probability described by Eq.~\eqref{eq:w_d}.

\subsubsection{1D qubit chain}

\begin{figure}[t] 
    \centering
    \includegraphics[width = 0.4\columnwidth]{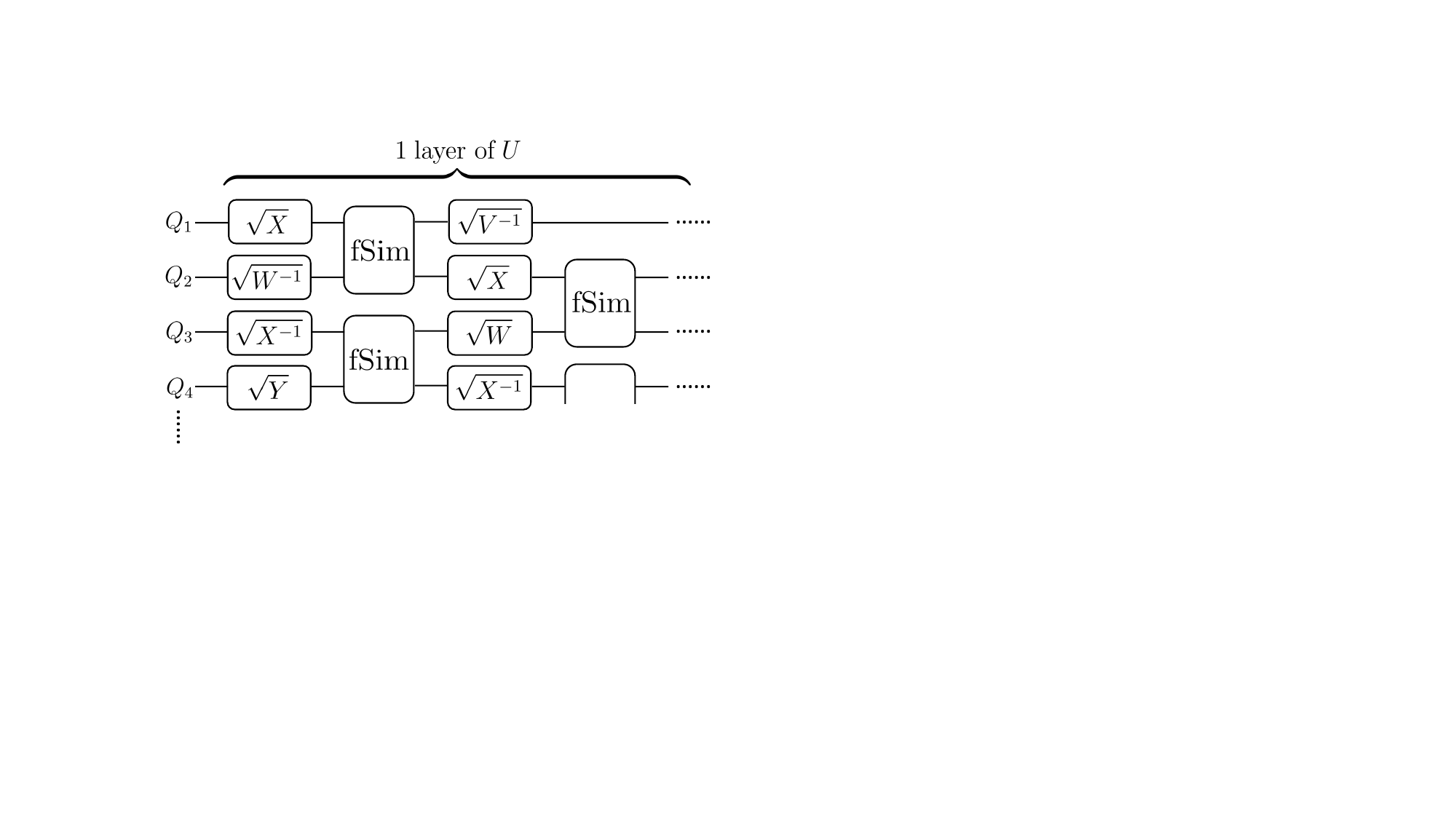}
    \caption{Quantum circuit representing one layer of operator $U$ in one-dimensional simulations. This layer consists of single-qubit gates (randomly chosen from $\sqrt{X^{\pm1}}$, $\sqrt{Y^{\pm1}}$, $\sqrt{W^{\pm1}}$, and $\sqrt{V^{\pm1}}$) and entangling gates $\text{fSim}(\theta,\phi,\beta)$ given by Eq.~\eqref{eq:fSim}. Here $W=(X+Y)/\sqrt{2}$, $V=(X-Y)/\sqrt{2}$, while the parameters for the $\text{fSim}(\theta,\phi,\beta)$ are chosen to be $\beta=0$, $\theta=\pi/6$, $\phi=2\pi/3$. The entire operator $U$ consists of $K$ such layers. Throughout the one-dimensional simulations open boundary conditions are assumed.
 }
    \label{fig:circuit_nonint}
\end{figure}

\begin{figure}[t] 
    \centering
    \includegraphics[width = 0.31\columnwidth]{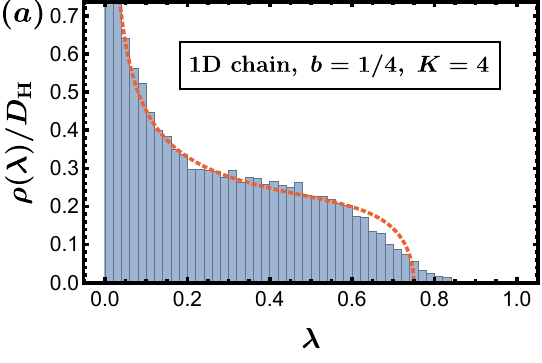}\quad
    \includegraphics[width = 0.31\columnwidth]{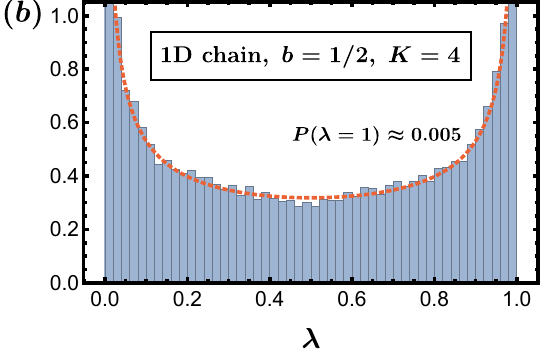}\quad
    \includegraphics[width = 0.31\columnwidth]{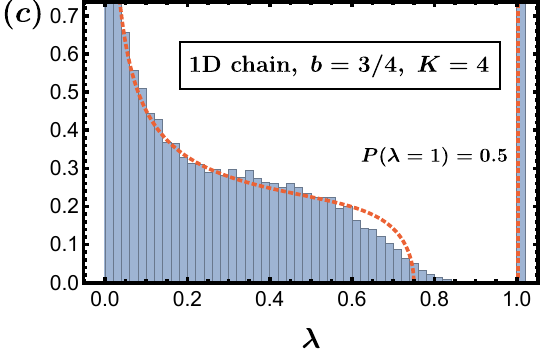}
 
    \vspace{0.2cm}
    
    \includegraphics[width = 0.31\columnwidth]{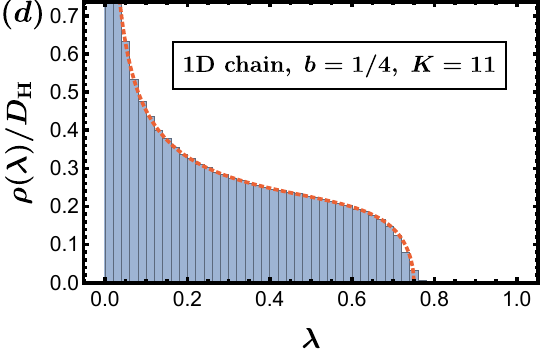}\quad
    \includegraphics[width = 0.31\columnwidth]{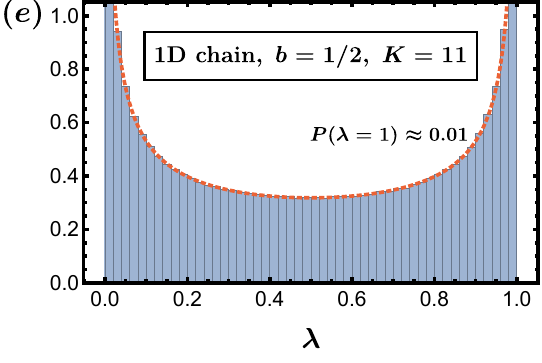}\quad
    \includegraphics[width = 0.31\columnwidth]{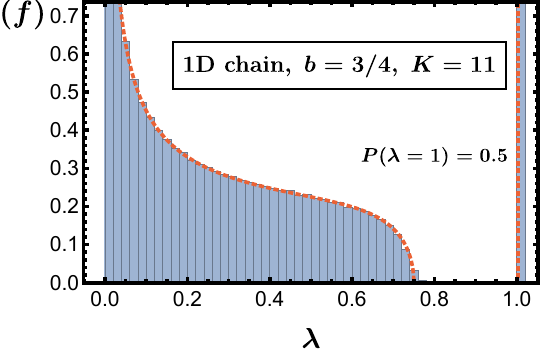}
    \caption{Histograms representing spectral density of a matrix $\Lambda_P$ of Eq.~\eqref{eq:Lambda_P} in a one-dimensional chain of $L=11$ qubits ($D_{\rm H}=2048$).
    Here the unitary $U$ entering $\Lambda_P$ consists of $K$ layers; an individual layer is depicted in Fig.~\ref{fig:circuit_nonint}.
    Averaging is performed over 2000 quantum circuit realizations for panels (a)-(c) and 100 realizations for panels (d)-(f).
    The density of $\lambda=1$ eigenvalues, $P(\lambda=1)$, is defined in Eq.~\eqref{eq:P_lambda_1}; here we take $\varepsilon=10^{-3}$.
    The red curve shows the analytical expression \eqref{eq:rho_lambda_math} for the spectral density of $\Lambda_P$ derived in Ref.~\cite{Collins2005} for the case when $U$ is Haar-random.
    }
    \label{fig:W_eval_hist_1D}
\end{figure}

Next, we present numerical results for one-dimensional qubit chain simulations.
Here we assume that the operator $U$ entering $\Lambda_P$ consists of $K$ layers, each layer is made of single-qubit and two-qubit gates as depicted in Fig.~\ref{fig:circuit_nonint}.
Single-qubit gates are drawn randomly from the collection of gates $\{\sqrt{X^{\pm1}}$, $\sqrt{Y^{\pm1}}$, $\sqrt{W^{\pm1}}$, $\sqrt{V^{\pm1}}\}$, where $X$, $Y$ are Paulis and $W=(X+Y)/\sqrt{2}$, $V=(X-Y)/\sqrt{2}$, while the two-qubit gates are the fSim gates:
\begin{equation}
	\text{fSim}(\theta,\phi,\beta) =
	\begin{pmatrix}
		1 & 0 & 0 & 0 \\
		0 & \cos\theta & i\exp(i\beta)\sin\theta & 0 \\
        0 & i\exp(-i\beta)\sin\theta & \cos\theta & 0 \\
        0 & 0 & 0 & \exp(i\phi)
	\end{pmatrix}
 \label{eq:fSim}
\end{equation}
with $\beta=0$, $\theta=\pi/6$, $\phi=2\pi/3$.
The projector $P$ measures $Q_1$ when $b=1/2$ and $Q_1, Q_2$ when $b=1/4$ or $b=3/4$.
Throughout the one-dimensional simulations open boundary conditions are assumed.
For this choice of parameters, the quantum circuit representing $U$ is non-integrable.

The results of the one-dimensional simulations are depicted in Fig.~\ref{fig:W_eval_hist_1D}.
Here we plot the histograms representing the spectrum of $\Lambda_P$ for the chain of $L=11$ qubits with $K=4$ (top row) and $K=L=11$ (bottom row).
For $K=4$ the basic features of the transition (depicted in Fig.~\ref{fig:W_eval_hist} for the random matrices) persist: for $b<1/2$ there are no eigenvalues at $\lambda=1$, while for $b>1/2$ there is a finite density of such eigenvalues.
This density is equal to $P(\lambda=1)=1/2$ for $b=3/4$.
The transition still takes place at $b=1/2$ where the gap closes.

Nevertheless, the continuum spectrum in the one-dimensional case for $K=4$ and $b<1/2$, $b>1/2$ possesses a tail that extends beyond the range of the random-matrix continuum.
That tail is not present when $U$ is comprised of $K=L=11$ layers, where the one-dimensional results coincide with the random-matrix results, see the bottom row of Fig.~\ref{fig:W_eval_hist_1D}.
This indicates that at that circuit depth the gate-generated unitary $U$ closely resembles a Haar-random unitary and is a manifestation of the fact that the quantum circuit comprising $U$ is chaotic.

\subsubsection{2D qubit array}

\begin{figure}[t] 
    \centering
    \includegraphics[width = 0.3\columnwidth]{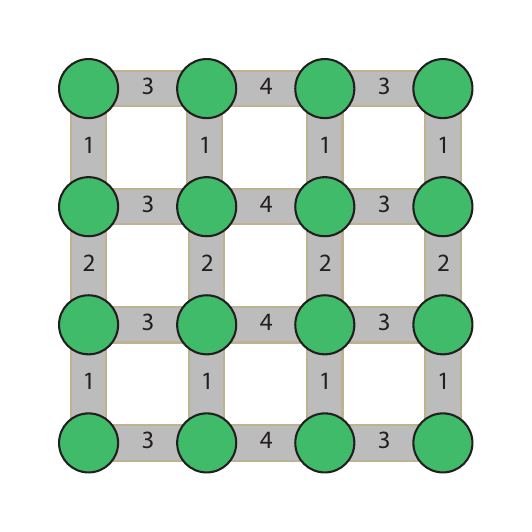}
    \caption{The order of operators for a two-dimensional grid of $N$ qubits.
    }
    \label{fig:Order}
\end{figure}

In two-dimensional simulations we consider the operator $U$ of the form $U=u^K$, where each individual unit operator $u$ contains four layers of XY rotations between the adjacent pairs of qubits: $u=\prod_i\text{fSim}_i(\theta=\pi/8,0,0)$. Here $\text{fSim}_i$  denotes an operator applied to the $i^\text{th}$ pair of qubits. The order of the pairs is indicated in Fig.~\ref{fig:Order}. For the simulations shown below we choose $K=20$. We have checked that the results are not affected if the depth of the circuit is increased to $K=100$ or the value of $\theta$ is changed. We set the projector $P$ by choosing three arbitrary qubits and projecting onto $M=1\ldots 7$ states of these three qubits. Changing the geometry of the qubit array or the location of the projection qubits affects the results only weakly. For the sizes available to us the results do not change drastically between consecutive values of $M$, so taking a finer step in $b$ would have little effect on the outcome. 

\begin{figure}[t] 
    \centering
    \includegraphics[width = 0.5\columnwidth]{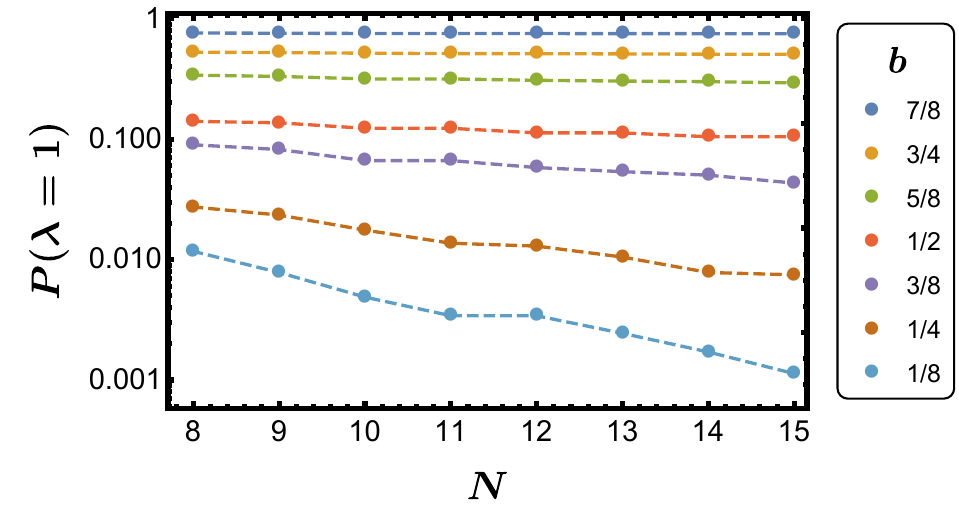}
    \caption{The density of eigenvalues at $\lambda=1$ of a matrix $\Lambda_P$ in a two-dimensional grid of $N$ qubits.
	}
    \label{fig:P_lambda}
\end{figure}

The main result of the simulations is the density of $\lambda=1$ eigenvalues, $P(\lambda=1)$, shown in Fig.~\ref{fig:P_lambda} as a function of the system size for various projection dimensions.
Figure~\ref{fig:P_lambda} clearly demonstrates that an exponential decrease of $P(\lambda=1)$ with $N$ for small $b$ is replaced by a constant at $b>1/2$, in agreement with the expectations (\ref{eq:Lambda_P_mx}). The transition occurs at $b_c\approx0.5-0.6$. Unfortunately, the available system sizes do not allow one to make a conclusion if the difference from $b_c=1/2$ (which we obtained for the random matrices and the one-dimensional model) is the finite size effect or not. 
\begin{figure}[t]
    \centering
    \includegraphics[width = 0.31\columnwidth]{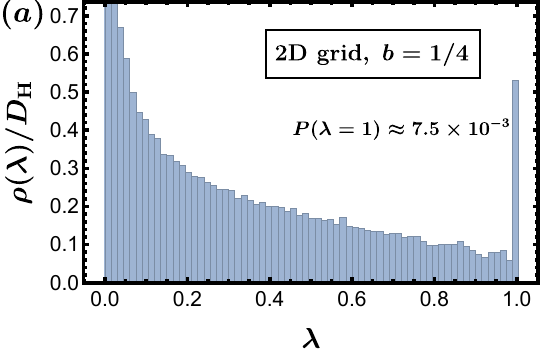}\quad
    \includegraphics[width = 0.31\columnwidth]{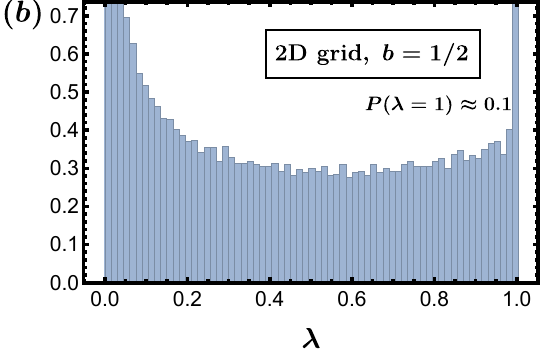}\quad
    \includegraphics[width = 0.31\columnwidth]{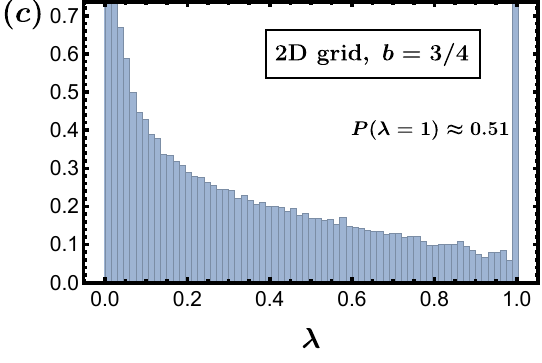}
    \caption{Histograms representing spectral density of a matrix $\Lambda_P$ of Eq.~\eqref{eq:Lambda_P} in a two-dimensional grid of $N=15$ qubits.
    The density of $\lambda=1$ eigenvalues, $P(\lambda=1)$, is defined in Eq.~\eqref{eq:P_lambda_1}; here we take $\varepsilon=10^{-3}$.
    }
    \label{fig:W_eval_hist_2D}
\end{figure}

The spectral density of $\Lambda_P$ for the two-dimensional case is shown in Fig.~\ref{fig:W_eval_hist_2D}. 
Whereas in the random unitary and in the one-dimensional evolution one observes a strong suppression of the density of states at large $\lambda$, here the suppression is not complete and the density of states remains finite (albeit small) even in the vicinity of $\lambda=1$. The available data does not allow us to distinguish whether it is a boundary-related effect or the phase transition in two dimensions occurs via a completely different mechanism from the one-dimensional and the random matrix case.
We leave this open question for a further study.

\section{\label{sec:RMT}Random matrix based solvable model}

\subsection{\label{sec:model_choice}The choice of the model}

Our goal is to calculate the average density of states 
\begin{equation}
    \rho(\lambda) = \left\langle\sum_{i=1}^{D_{\rm H}} \delta(\lambda-\lambda_i)\right\rangle = -\frac{1}{\pi}\text{Im}\left\langle\Tr[G(\lambda+i0^+)]\right\rangle,
    \label{eq:dos}
\end{equation}
where the Green function is given by
\begin{equation}
    G(\lambda) = \frac{1}{\lambda\openone-\Lambda_P}.
    \label{eq:G_init}
\end{equation}
The function $G(\lambda)$ written in this way is a many-body Green function, i.e. a $D_{\rm H}\times D_{\rm H}$ matrix (not to be confused with a single-particle Green function appearing in usual many-body calculations).
Furthermore, $G(\lambda)$ of Eq.~\eqref{eq:G_init} is an analytic function except for the line $\text{Im}(\lambda)=0$, $0\leq\text{Re}(\lambda)\leq 1$, where it has poles.

The nontrivial procedure here is to introduce the averaging $\langle\dots\rangle$ in Eq.~\eqref{eq:dos}.
If $\Lambda_P$ were simple Gaussian random matrices, it would have been possible to perform the averaging, especially in the interesting $D_{\rm H}\gg 1$ limit, using well-established diagrammatic techniques~\cite{AGD}.
However, the structure of the operator $\Lambda_P$ of Eq.~\eqref{eq:Lambda_P} requires an augmentation of these techniques, which we describe below.

Let us introduce $D_{\rm H}\times D_{\rm H}$ matrix $H$ from the Gaussian unitary ensemble (GUE):
\begin{subequations}\label{eq:GUE}
\begin{align}
&\langle H_{ij}\rangle=0, \\
&\langle H_{ij} H_{kl}\rangle=\frac{\gamma}{D_{\rm H}} \delta_{il} \delta_{jk}, \quad i,j,k,l=1, \dots, D_{\rm H}, \label{eq:H_ijHkl}
\end{align}
\end{subequations}
where parameter $\gamma$ is introduced to formally control the perturbation series; the main features of the results will not depend on $\gamma$.
The unitary evolution $U$ is modeled by the Cayley transform of the GUE (the advantage of such representation will become clear shortly):
\begin{equation}
    U = \frac{\openone-iH}{\openone+iH}.
    \label{eq:cayley}
\end{equation}
The matrix $U$ written in this way is unitary; its eigenfunctions are the eigenfunctions of $H$, and therefore they obey the Porter-Thomas distribution.
In addition, it is known~\cite{Brouwer1995, Brouwer&Beenakker96} that for $\gamma=2$, the ensemble averaged $\langle U \rangle=0$, thus mimicking the circular unitary ensemble (CUE) distribution of the eigenvalue density (even though the eigenvalue repulsion here is different from the CUE case, it will not lead to any changes in the results). 

One could use Eq.~\eqref{eq:Lambda_P} with $U$ given by Eq.~\eqref{eq:cayley} in Eqs.~\eqref{eq:dos}-\eqref{eq:G_init} and try to perform the averaging.
However, in this case $\Lambda_P$ is a non-linear function of $H$ and the calculations are difficult. In order to simplify them we note that the denominator in Eq.~\eqref{eq:cayley} appears when the matrix
\begin{align}
	& \check L =
	\begin{pmatrix}
		\openone & \openone \\
		H & i \openone
	\end{pmatrix},
 \label{eq:L}
\end{align}
which is linear in $H$, is inverted. Indeed, 
\begin{align}
	U=\begin{pmatrix}
	\openone; & -\openone	
	\end{pmatrix}
	\check L^{-1}
	\begin{pmatrix}
	\openone \\
	0	
	\end{pmatrix}.
\end{align}
In the following we shall indicate the structures (vectors and matrices) in this auxiliary subspace by a subscript A. 
Unfortunately, the matrix (\ref{eq:L}) is not Hermitian. To deal with it we need to double the space of the problem again, which consists of ``forward" and ``backward" evolution:
\begin{align}
	& 
	\begin{pmatrix}
		0 & \check L \\
		\check L^{\dagger} & 0
	\end{pmatrix}.
\end{align}
We shall indicate the vectors and matrices in this subspace by a subscript K. This allows us to represent the computation of the Green function of the original problem as a computation of the Green function in the extended space of $4D_{\rm H}\times4D_{\rm H}$ dimensions. The enlarged Hilbert space is a product:
\begin{equation}
    \mathbb{H}^{(4D_{\rm H}\times4D_{\rm H})} = \mathbb{H}^{(2\times2)}_{\rm K}\otimes\mathbb{H}^{(2\times2)}_{\rm A}\otimes\mathbb{H}^{(D_{\rm H}\times D_{\rm H})}_{\rm H},
\end{equation}
where the subscript $\rm H$ denotes the original Hilbert space.
We start our calculation by defining the matrix in the $\mathbb{H}^{(4D_{\rm H}\times4D_{\rm H})}$ space:
\begin{equation}
	\hat{\mathcal{M}}(\xi) =
	\begin{pmatrix}
		\check 0 & \check L \\
		\check L^{\dagger} & \check 0
	\end{pmatrix}_{\rm K}
	+\sqrt{\xi}
	\begin{pmatrix}
		\check t_1 & \check 0 \\
		\check 0 & \check t_2
	\end{pmatrix}_{\rm K}
	\equiv\hat{\mathcal{L}}+\sqrt{\xi}\hat{\mathcal{T}}, \label{eq:M_lambda_def}
\end{equation}
where $\check{t}_{1,2}$ are matrices in the $\mathbb{H}_{\rm A}\otimes\mathbb{H}_{\rm H}$ space and $\xi$ is a complex number.
Straightforward calculation (see \ref{sec:App_Eq_27}) yields
\begin{align}
    -\frac{\sqrt{\xi}}{2}\Tr_{\rm K,\rm A,\rm H}[\hat{\mathcal{T}}\hat{\mathcal{M}}^{-1}(\xi)]=\sum_{k=1}^{\infty}\xi^{k}\Tr_{\rm A,\rm H}\left[(\check L^{\dagger})^{-1}\check t_2\check L^{-1}\check t_1\right]^k.
    \label{eq:series1}
\end{align}
Comparing the series in Eq.~\eqref{eq:series1} with a formal expansion of the Green function $G(\lambda)$ of Eq.~\eqref{eq:G_init} in the powers of $1/\lambda$,
\begin{equation}
	\Tr_{\rm H}[G(\lambda)]= \sum_{k=0}^\infty \frac{1}{\lambda^{k+1}}\Tr_{\rm H}[\Lambda_P^k],
	\label{eq:series2}
\end{equation}
we note that they become equivalent if 
\begin{equation}
    \Tr_{\rm A,\rm H}\left[(\check L^{\dagger})^{-1}\check t_2\check L^{-1}\check t_1\right]^k = \Tr_{\rm H}[\Lambda_P^k].
    \label{eq:Lt2Lt1}
\end{equation}
We now choose $\check t_{1,2}$ so that Eq.~\eqref{eq:Lt2Lt1} is satisfied:
\begin{align}
	\check t_1=2P
	\begin{pmatrix}
		\openone & 0 \\
		0 & 0
	\end{pmatrix}_{\rm A}, \quad
	\check t_2=\frac{P}{2}
	\begin{pmatrix}
		\openone & -\openone \\
		-\openone & \openone
	\end{pmatrix}_{\rm A}, \label{eq:t12}
\end{align}
which yields
\begin{equation}
    P(\check L^{\dagger})^{-1}\check t_2\check L^{-1}\check t_1 = \Lambda_P
    \begin{pmatrix}
		\openone & 0 \\
		0 & 0
	\end{pmatrix}_{\rm A}.
\end{equation}
Equation \eqref{eq:Lt2Lt1} then clearly follows because $\check t_1P = \check t_1$.
As a result, using Eqs.~\eqref{eq:series1}-\eqref{eq:Lt2Lt1} we obtain
\begin{equation}
	\Tr_{\rm H}[G(\lambda)]= \frac{1}{\lambda}\left\{D_{\rm H}-\frac{1}{2\sqrt{\lambda}}\Tr_{\rm K,\rm A,\rm H}[\hat{\mathcal{T}}\hat{\mathcal{M}}^{-1}(1/\lambda)]\right\},
	\label{eq:G_M}
\end{equation}
where $\xi=1/\lambda$.
Equations~\eqref{eq:L},\eqref{eq:M_lambda_def},\eqref{eq:t12} and \eqref{eq:G_M} constitute the main results of this subsection.
We succeeded in reformulating the task of computing $\left\langle\Tr_{\rm H}[G(\lambda)]\right\rangle$ in terms of inverting the matrix $\hat{\mathcal{M}}(\xi)$ linear in disorder and performing its ensemble-averaging $\langle \hat{\mathcal{M}}^{-1}(\xi) \rangle$. 
The diagrammatic techniques for conducting this are well-known and we discuss their application to our model in the  following subsection.

\subsection{Diagrammatic calculation}

We now move on to performing the averaging ${\hat{\mathcal{G}}(\xi)=\langle \hat{\mathcal{M}}^{-1}(\xi) \rangle}$.
To this end, we write the matrix $\hat{\mathcal{M}}$ as:
\begin{equation}
	\hat{\mathcal{M}}(\xi)=\hat{\mathcal{M}}_0(\xi)+H\Pi_x,
\end{equation}
where we have introduced $\hat{\mathcal{M}}_0\equiv\hat{\mathcal{M}}|_{H=0}$ and
\begin{equation}
	\Pi_x=
	\begin{pmatrix}
		0 & 0 & 0 & 0 \\
		0 & 0 & 1 & 0 \\
		0 & 1 & 0 & 0 \\
		0 & 0 & 0 & 0 \\
	\end{pmatrix}_{\rm K,A}.
	\label{eq:pi_x}
\end{equation}
Hence,
\begin{equation}
	\hat{\mathcal{G}}(\xi)=\left\langle\frac{1}{\hat{\mathcal{G}}^{-1}_0(\xi)+H\Pi_x} \right\rangle,
	\label{eq:G_def}
\end{equation}
where $\hat{\mathcal{G}}^{-1}_0\equiv\hat{\mathcal{M}}_0$.
To proceed with the diagrammatic calculation, we introduce diagrams for the bare and the dressed Green functions,
\begin{equation}
\begin{tikzpicture}
  \begin{feynman}
    \vertex (i);
    \vertex [right=1.2cm of i] (o);
    \diagram*{
      (o) --[fermion] (i)      
    };
  \end{feynman}
\end{tikzpicture}
~\equiv~\hat{\mathcal{G}}_0,\quad
\begin{tikzpicture}
  \begin{feynman}
    \vertex (i);
    \vertex [right=1.2cm of i] (o);
    \diagram*{
      (o) --[fermion, line width=0.6mm] (i)      
    };
  \end{feynman}
\end{tikzpicture}
~\equiv~\langle\hat{\mathcal{G}}\rangle,
\end{equation}
as well as for performing the averaging \eqref{eq:H_ijHkl}:
\begin{equation}
\begin{tikzpicture}
  \begin{feynman}
    \vertex (i1);
    \node[right=0.4cm of i1,draw,fill=white,circle] (v1){$H$};
    \vertex [right=0.8cm of v1] (o1);
    \vertex [right=0.8cm of o1] (i2);
    \node[right=0.4cm of i2,draw,fill=white,circle] (v2){$H$};
    \vertex [right=0.8cm of v2] (o2);
    \diagram*{
      (v1) --[fermion, edge label=$i$] (i1),        
      (o1) --[fermion, edge label=$j$]  (v1)
    };
    \diagram*{
      (v2) --[fermion, edge label=$k$] (i2),        
      (o2) --[fermion, edge label=$l$]  (v2),
      (v2) --[scalar, half right,edge label=$\frac{\gamma}{D_{\rm H}}\delta_{il}\delta_{jk}$] (v1)
    };
  \end{feynman}
\end{tikzpicture}~.
\end{equation}
The usual Dyson's equation in the diagrammatic form then reads:
\begin{equation}
\vcenter{\hbox{\begin{tikzpicture}
  \begin{feynman}
    \vertex (i);
    \vertex [right=1.2cm of i] (o);
    \diagram*{
      (o) --[fermion, line width=0.6mm] (i)      
    };
  \end{feynman}
\end{tikzpicture}}}
~=~
\vcenter{\hbox{\begin{tikzpicture}
  \begin{feynman}
    \vertex (i);
    \vertex [right=1.2cm of i] (o);
    \diagram*{
      (o) --[fermion] (i)     
    };
  \end{feynman}
\end{tikzpicture}}}
~+~
\vcenter{\hbox{\begin{tikzpicture}
  \begin{feynman}
    \vertex (i);
    \node[right=1.2cm of i,draw,fill=white,circle] (v){$\hat{\varSigma}$};
    \vertex [right=1.45cm of v] (o);
    \diagram*{
      (v) --[fermion] (i),        
      (o) --[fermion, line width=0.6mm]  (v)
    };
  \end{feynman}
\end{tikzpicture}}}~,\label{eq:Dyson}
\end{equation}
where $\hat{\varSigma}$ is the self-energy.
The expansion of $\hat{\varSigma}$ in powers of $H$ can then be written in the following diagrammatic form:
\newcommand{\tallvdots}{%
  \vcenter{%
    \baselineskip=4pt \lineskiplimit=0pt
    \hbox{.}\hbox{.}\hbox{.}
    \hbox{.}\hbox{.}\hbox{.}
    \hbox{.}\hbox{.}\hbox{.}
    \hbox{.}\hbox{.}\hbox{.}
    \hbox{.}\hbox{.}\hbox{.}
  }%
}
\begin{alignat}{2}
\vcenter{\hbox{\begin{tikzpicture}
  \begin{feynman}
    \node[right=1.5cm of i,draw,fill=white,circle] (v){$\hat\varSigma$};
    \diagram*{
      (v)
    };
  \end{feynman}
\end{tikzpicture}}}
~&=~
\begin{tikzpicture}
  \begin{feynman}
  \tikzfeynmanset{every vertex={empty dot}}
    \vertex (i);
    \vertex [right=1.5cm of i] (o);
    \diagram*{
      (o) --[fermion] (i);    
    };
    \diagram*{
      (i) --[scalar, half left,edge label=$\propto\gamma$] (o)
    };
  \end{feynman}
\end{tikzpicture}
~~+~~
\begin{tikzpicture}
  \begin{feynman}
  \tikzfeynmanset{every vertex={empty dot}}
    \vertex (i);
    \vertex [right=1cm of i, empty dot] (v1);
    \vertex [right=1cm of v1, empty dot] (v2);
    \vertex [right=1cm of v2, empty dot] (o);
    \diagram*{
      (o) --[fermion] (v2)  --[fermion] (v1)  --[fermion] (i)  
    };
    \diagram*{
      (i) --[scalar, half left,edge label=$\quad\quad\propto\gamma^2$] (o)
    };
    \diagram*{
      (v1) --[scalar, half left] (v2)
    };
  \end{feynman}
\end{tikzpicture}
~~+&&\mathrel{\tallvdots}~
\begin{tikzpicture}
  \begin{feynman}
  \tikzfeynmanset{every vertex={empty dot}}
    \vertex (i);
    \vertex [right=1cm of i, empty dot] (v1);
    \vertex [right=1cm of v1, empty dot] (v2);
    \vertex [right=1cm of v2, empty dot] (o);
    \diagram*{
      (o) --[fermion] (v2)  --[fermion] (v1)  --[fermion] (i)  
    };
    \diagram*{
      (i) --[scalar, half left] (v2)
    };
    \diagram*{
      (v1) --[scalar, half left,edge label=$\propto\gamma^2/D_{\rm H}$] (o)
    };
  \end{feynman}
\end{tikzpicture}
~+~\nonumber\\
~&+~
\begin{tikzpicture}
  \begin{feynman}
  \tikzfeynmanset{every vertex={empty dot}}
    \vertex (i);
    \vertex [right=0.5cm of i, empty dot] (v1);
    \vertex [right=0.5cm of v1, empty dot] (v2);
    \vertex [right=0.5cm of v2, empty dot] (v3);
    \vertex [right=0.5cm of v3, empty dot] (v4);
    \vertex [right=0.5cm of v4, empty dot] (o);
    \diagram*{
      (o) --[fermion] (v4) --[fermion] (v3) --[fermion] (v2)  --[fermion] (v1)  --[fermion] (i)  
    };
    \diagram*{
      (i) --[scalar, half left,edge label=$\quad\quad\propto\gamma^3$] (o)
    };
    \diagram*{
      (v1) --[scalar, half left] (v2)
    };
    \diagram*{
      (v3) --[scalar, half left] (v4)
    };
  \end{feynman}
\end{tikzpicture}
~+~
\begin{tikzpicture}
  \begin{feynman}
  \tikzfeynmanset{every vertex={empty dot}}
    \vertex (i);
    \vertex [right=0.5cm of i, empty dot] (v1);
    \vertex [right=0.5cm of v1, empty dot] (v2);
    \vertex [right=0.5cm of v2, empty dot] (v3);
    \vertex [right=0.5cm of v3, empty dot] (v4);
    \vertex [right=0.5cm of v4, empty dot] (o);
    \diagram*{
      (o) --[fermion] (v4) --[fermion] (v3) --[fermion] (v2)  --[fermion] (v1)  --[fermion] (i)  
    };
    \diagram*{
      (i) --[scalar, half left,edge label=$\quad\quad\propto\gamma^3$] (o)
    };
    \diagram*{
      (v1) --[scalar, half left] (v4)
    };
    \diagram*{
      (v2) --[scalar, half left] (v3)
    };
  \end{feynman}
\end{tikzpicture}
~+&&\mathrel{\tallvdots}~
\begin{tikzpicture}
  \begin{feynman}
  \tikzfeynmanset{every vertex={empty dot}}
    \vertex (i);
    \vertex [right=0.5cm of i, empty dot] (v1);
    \vertex [right=0.5cm of v1, empty dot] (v2);
    \vertex [right=0.5cm of v2, empty dot] (v3);
    \vertex [right=0.5cm of v3, empty dot] (v4);
    \vertex [right=0.5cm of v4, empty dot] (o);
    \diagram*{
      (o) --[fermion] (v4) --[fermion] (v3) --[fermion] (v2)  --[fermion] (v1)  --[fermion] (i)  
    };
    \diagram*{
      (i) --[scalar, half left,edge label=$\quad\quad\propto\gamma^3/D_{\rm H}$] (v4)
    };
    \diagram*{
      (v1) --[scalar, half left] (v2)
    };
    \diagram*{
      (v3) --[scalar, half left] (o)
    };
  \end{feynman}
\end{tikzpicture}
~+~
\begin{tikzpicture}
  \begin{feynman}
  \tikzfeynmanset{every vertex={empty dot}}
    \vertex (i);
    \vertex [right=0.5cm of i, empty dot] (v1);
    \vertex [right=0.5cm of v1, empty dot] (v2);
    \vertex [right=0.5cm of v2, empty dot] (v3);
    \vertex [right=0.5cm of v3, empty dot] (v4);
    \vertex [right=0.5cm of v4, empty dot] (o);
    \diagram*{
      (o) --[fermion] (v4) --[fermion] (v3) --[fermion] (v2)  --[fermion] (v1)  --[fermion] (i)  
    };
    \diagram*{
      (i) --[scalar, half left,edge label=$\quad\quad\propto\gamma^3/D_{\rm H}$] (v4)
    };
    \diagram*{
      (v1) --[scalar, half left] (o)
    };
    \diagram*{
      (v2) --[scalar, half left] (v3)
    };
  \end{feynman}
\end{tikzpicture}
~+\dots~\label{eq:SCBA_diag1}\\
~&=~
\begin{tikzpicture}
  \begin{feynman}
  \tikzfeynmanset{every vertex={empty dot}}
    \vertex (i);
    \vertex [right=1.5cm of i] (o);
    \diagram*{
      (o) --[fermion, line width=0.6mm] (i);    
    };
    \diagram*{
      (i) --[scalar, half left] (o)
    };
  \end{feynman}
\end{tikzpicture}~,\label{eq:SCBA_diag2}
\end{alignat}
where for simplicity we have denoted 
\begin{equation}
\vcenter{\hbox{\begin{tikzpicture}
  \begin{feynman}
    \node[right=0.4cm of i1,draw,fill=white,circle] (v1){$H$};
  \end{feynman}
\end{tikzpicture}}}
~\equiv~
\begin{tikzpicture}
  \begin{feynman}
  \tikzfeynmanset{every vertex={empty dot}}
    \vertex (i);
  \end{feynman}
\end{tikzpicture}~.
\end{equation}
All crossing diagrams in the expansion~\eqref{eq:SCBA_diag1} are parametrically small in $1/D_{\rm H}$, leaving only non-crossing diagrams in the expansion. 
Resummation of the non-crossing diagrams yields Eq.~\eqref{eq:SCBA_diag2}.
This is known as the self-consistent Born approximation (SCBA)~\cite{AGD} and it is exact in the thermodynamic limit $D_{\rm H}\to\infty$.
Equations~\eqref{eq:Dyson},\eqref{eq:SCBA_diag2} together form a system of equations known as the ``self-consistency equations", and we move on to their analysis in the next subsection.

\subsection{Self-consistency equations}

Using diagrammatic techniques we obtained the following self-consistency equations~\eqref{eq:Dyson},\eqref{eq:SCBA_diag2} in the limit $D_{\rm H}\to\infty$:
\begin{subequations}
\begin{align}
&\hat{\mathcal{G}}=\hat{\mathcal{G}}_0+\hat{\mathcal{G}}_0\hat\varSigma\hat{\mathcal{G}}, \label{eq:g_g0_sigma2}\\
	&\hat\varSigma=\openone\otimes\frac{\gamma}{D_{\rm H}}\Pi_x \Tr_{\rm H}[\hat{\mathcal{G}}]\Pi_x.\label{eq:sigma_scborn}
\end{align}
\end{subequations}
Note that since $ \Tr_{\rm H}[\hat{\mathcal{G}}]\propto D_{\rm H}, $ the solution of this system is independent of $D_{\rm H}$.
This system of equations is closed, but it is non-linear and is written in the form of $4\times 4$ matrices.
However, it can still can be solved; we describe the procedure to solve it below.

We start by introducing a projector onto the inner $2\times2$ subspace of the $\mathbb H_{\rm K}\otimes\mathbb H_{\rm A}$, which we denote $\mathbb H_{\rm I}$:
\begin{equation}
	\hat\Pi=
	\begin{pmatrix}
		0 & 0 & 0 & 0 \\
		0 & 1 & 0 & 0 \\
		0 & 0 & 1 & 0 \\
		0 & 0 & 0 & 0 \\
	\end{pmatrix}_{\rm K,A},
    \label{eq:Pi}
\end{equation}
Next, using the fact that $\hat\Pi_x\hat\Pi = \hat\Pi\hat\Pi_x = \hat\Pi_x$, and multiplying Eq.~\eqref{eq:g_g0_sigma2} by the projector $\hat \Pi$ from both sides, we rewrite it as
\begin{equation}
    g^{-1}=g_0^{-1}-\varsigma,
    \label{eq:g_g0_sigma3}
\end{equation}
where $\varsigma$, $g$ and $g_0$ are the inner blocks of the matrices $\hat\varSigma$, $\hat{\mathcal{G}}$ and $\hat{\mathcal{G}}_0$, respectively:
\begin{equation}
	\hat\varSigma=
	\begin{pmatrix}
		0 & 0 & 0 & 0 \\
		0 & \varsigma_{11} & \varsigma_{12} & 0 \\
		0 & \varsigma_{21} & \varsigma_{22} & 0 \\
		0 & 0 & 0 & 0 \\
	\end{pmatrix}_{\rm K,A}, \quad
    \hat\Pi \hat{\mathcal{G}} \hat\Pi=
	\begin{pmatrix}
		0 & 0 & 0 & 0 \\
		0 & g_{11} & g_{12} & 0 \\
		0 & g_{21} & g_{22} & 0 \\
		0 & 0 & 0 & 0 \\
	\end{pmatrix}_{\rm K,A},
\end{equation}
and similarly for $\hat\Pi \hat{\mathcal{G}}_0 \hat\Pi$.
Direct calculation (see \ref{sec:App_Eq_42}) then gives
\begin{equation}
	g_0^{-1}=P(f_2\tau_2 -f_0 \tau_0)+\bar P\tau_2,
	\label{eq:g0m1}
\end{equation}
where $\bar P = \openone-P$, $\{\tau_i\}_{i=0}^4$ are the standard Pauli matrices in the space $\mathbb H_{\rm I}$, and
\begin{align}
	f_2=\frac{1+\xi}{1-\xi},\quad f_0=-\frac{2\sqrt{\xi}}{1-\xi}.
    \label{eq:f2f0}
\end{align}
Note that
\begin{equation}
	f_2^2-f_0^2=1.
	\label{eq:fy_f1_1}
\end{equation}
To proceed, we assume that $\varsigma$ has the same form as $g_0^{-1}$:
\begin{equation}
	\varsigma=(u\tau_0-v\tau_2)\otimes\openone.
	\label{eq:varsigma}
\end{equation}
Then, substituting Eqs.~\eqref{eq:g0m1},\eqref{eq:varsigma} into Eq.~\eqref{eq:g_g0_sigma3} gives
\begin{equation}
    g^{-1}=P\left[(f_2+v)\tau_2 -(f_0+u) \tau_0\right]+\bar{P}\left[(1+v)\tau_2-u\tau_0\right].
    \label{eq:g_-1_1}
\end{equation}
Inverting $g^{-1}$ of Eq.~\eqref{eq:g_-1_1} then yields 
\begin{equation}
	g= P \frac{(f_0+u)\tau_0+(f_2+v)\tau_2}{(f_2+v)^2-(f_0+u)^2}+\bar P \frac{u\tau_0+(1+v)\tau_2}{(1+v)^2-u^2}.
	\label{eq:g_g1_g2}
\end{equation}
At the same time, Eq.~\eqref{eq:sigma_scborn} in terms of $\varsigma$, $g$ becomes
\begin{align}
	\varsigma=\openone\otimes\frac{\gamma}{D_{\rm H}}\tau_1 \Tr_{\rm H}[g]\tau_1.
    \label{eq:varsigma_2}
\end{align}
Substituting Eq.~\eqref{eq:g_g1_g2} together with Eq.~\eqref{eq:varsigma} into Eq.~\eqref{eq:varsigma_2}, we obtain the self-consistency equations:
\begin{align}
\begin{cases}
	\dfrac{u}{\gamma}=\dfrac{b(f_0+u)}{(f_2+v)^2-(f_0+u)^2}+\dfrac{\bar{b} u}{(1+v)^2-u^2},\\
	\dfrac{v}{\gamma}=\dfrac{b(f_2+v)}{(f_2+v)^2-(f_0+u)^2}+\dfrac{\bar{b}(1+v)}{(1+v)^2-u^2},
\end{cases}\label{eq:sc_uv12}
\end{align}
where $\Tr[P] = bD_{\rm H}$ and $\bar b = 1-b$.
The system \eqref{eq:sc_uv12} can be simplified using a new set of variables:
\begin{equation}
	v_{\pm}=v\pm u, \quad f_{\pm}=f_2\pm f_0=\frac{1\mp\sqrt\xi}{1\pm\sqrt\xi},
\end{equation}
such that it becomes
\begin{align}
&\begin{cases}
	\dfrac{v_+}{\gamma}=\dfrac{\bar b}{v_-+1}+\dfrac{b}{v_-+f_-}, \\
	\dfrac{v_-}{\gamma}=\dfrac{\bar b}{v_++1}+\dfrac{b}{v_++f_+}. \\
\end{cases}\label{eq:sc_vpm}
\end{align}
Equations \eqref{eq:varsigma} and \eqref{eq:sc_vpm} are the main results of this subsection: they allow one to obtain the expression for the self-energy in the limit $D_{\rm H}\gg 1$.
Before embarking onto the analysis of the non-linear system \eqref{eq:sc_vpm}, we derive the expression for the density of states in terms of $v_{\pm}$.

\subsection{Density of states via solution of the self-consistency equations}

To obtain the Green function of Eq.~\eqref{eq:G_M}, we need to compute
\begin{align}
    \left\langle\Tr_{\rm K,\rm A,\rm H}[\hat{\mathcal{T}}\hat{\mathcal{M}}^{-1}(\xi)]\right\rangle \equiv \Tr_{\rm K,\rm A,\rm H}[\hat{\mathcal{T}}\hat{\mathcal{G}}(\xi)] = \Tr_{\rm K,\rm A,\rm H}\left[\hat{\mathcal{T}}\hat{\mathcal{G}}_0\right] + \Tr_{\rm K,\rm A,\rm H}\left[\hat{\mathcal{G}}_0\hat{\mathcal{T}}\hat{\mathcal{G}}_0\hat\varSigma(\hat\openone-\hat{\mathcal{G}}_0\hat\varSigma)^{-1}\right],
    \label{eq:Tr_KAH_1}
\end{align}
where $\hat\openone$ is the identity matrix in the $\mathbb{H}^{(4D_{\rm H}\times4D_{\rm H})}$ space.
Using the fact that $\hat\varSigma\hat\Pi = \hat\Pi\hat\varSigma = \hat\varSigma$, we can project Eq.~\eqref{eq:Tr_KAH_1} onto the $\mathbb H_{\rm I}$ space, obtaining:
\begin{equation}
    \Tr_{\rm K,\rm A,\rm H}[\hat{\mathcal{T}}\hat{\mathcal{G}}(\xi)] = \Tr_{\rm K,\rm A,\rm H}\left[\hat{\mathcal{T}}\hat{\mathcal{G}}_0\right] + \Tr_{\rm I, \rm H}\left[ \eta \varsigma g g_0^{-1}\right],
    \label{eq:Tr_T_G}
\end{equation}
where
\begin{equation}
    \hat\Pi \hat{\mathcal{G}}_0\hat{\mathcal{T}}\hat{\mathcal{G}}_0 \hat\Pi=
	\begin{pmatrix}
		0 & 0 & 0 & 0 \\
		0 & \eta_{11} & \eta_{12} & 0 \\
		0 & \eta_{21} & \eta_{22} & 0 \\
		0 & 0 & 0 & 0 \\
	\end{pmatrix}_{\rm K,A}.
\end{equation}
Straightforward calculation (see \ref{sec:App_Eq_55_56}) yields:
\begin{equation}
	\Tr_{\rm K,\rm A,\rm H}\left[\hat{\mathcal{T}}\hat{\mathcal{G}}_0\right]= f_0 b D_{\rm H}
	\label{eq:tr_g0_t}
\end{equation}
and
\begin{equation}
	\eta=\frac{2P}{1-\xi}\left(f_0\tau_2 +f_2\tau_0\right).
	\label{eq:pi_g0_t_g0_pi}
\end{equation}
Using Eqs.~\eqref{eq:g0m1}, \eqref{eq:varsigma}, \eqref{eq:g_g1_g2}, \eqref{eq:tr_g0_t} and \eqref{eq:pi_g0_t_g0_pi} in Eq.~\eqref{eq:Tr_T_G}, and further substituting the result into Eq.~\eqref{eq:G_M} while replacing $\xi\to 1/\lambda$, we obtain
\begin{equation}
	\Tr_{\rm H}[G(\lambda)]= \frac{D_{\rm H}}{\lambda}\left\{1+\frac{b}{\lambda-1}\left[1-\frac{4\bar b\lambda}{(\lambda-1)\mathcal{F}}\right]\right\},
    \label{eq:Tr_G_pm}
\end{equation}
where 
\begin{align}
	\mathcal{F}=\gamma^{-1}(v_-+1)(v_++1)(v_++f_+)(v_-+f_-)-\bar b(v_++f_+)(v_-+f_-)-b(v_-+1)(v_++1)
	\label{eq:F}
\end{align}
and
\begin{align}
	f_{\pm}=\frac{\sqrt\lambda\mp1}{\sqrt\lambda\pm1}.
    \label{eq:f_pm}
\end{align}
Here $v_{\pm}$ are the solutions of the self-consistency equations \eqref{eq:sc_vpm}.
Equations~\eqref{eq:sc_vpm}, \eqref{eq:Tr_G_pm}-\eqref{eq:f_pm} constitute the complete solution for the density of states \eqref{eq:dos}.

\subsection{Analysis of the self-consistency equations \eqref{eq:sc_vpm} for $|\lambda|>1$}

In the region $|\lambda|>1$ the function $\Tr_{\rm H}[G(\lambda)]$ is analytical and is equal to its Laurent series.
To check the consistency of this statement, we perturbatively solve for $\Tr_{\rm H}[G(\lambda)]$ in the limit $\lambda\to\infty$.
We start by noting that the system \eqref{eq:sc_vpm} and the Green function \eqref{eq:Tr_G_pm}-\eqref{eq:F} are invariant under a symmetry transformation $\sqrt{\lambda}\to - \sqrt{\lambda}$, $f_+\leftrightarrow f_-$, and $v_+\leftrightarrow v_-$.
Hence, the expansion of $\Tr_{\rm H}[G(\lambda)]$ should feature only even powers of $\sqrt{\lambda}$:
\begin{equation}
	\Tr_{\rm H}[G(\lambda)]= D_{\rm H}\sum_{k=0}^{\infty}\frac{a_k}{\lambda^{k+1}}.
 \label{eq:Tr_G_expan}
\end{equation}
The coefficients $a_i$ can be computed by solving Eqs.~\eqref{eq:sc_vpm} iteratively.
In the first iteration we set $f_{\pm}=1$ and after straightforward calculation (see \ref{sec:App_Eq_61}) we obtain:
\begin{equation}
    a_0 = 1,\ a_1 = b\left[1-\frac{32\gamma\bar b}{\left(1+\sqrt{1+4\gamma}\right)^3} \right].
    \label{eq:a0a1}
\end{equation}
In fact, coefficients $a_k$ are nothing but the moments of the density of states:
\begin{equation}
    \int_{0}^1 \lambda^k \rho(\lambda)d\lambda = \oint \frac{d\lambda}{2\pi i} \Tr_{\rm H}[G(\lambda)]\lambda^k = D_{\rm H} a_k,
\end{equation}
where the second integral is taken over any contour encompassing the cut $0\leq \text{Re}[\lambda]\leq1$, $\text{Im}(\lambda)=0$.

Here $a_0=1$ signifies the correct normalization of the density of states and is model independent.
On the other hand, $a_1$ does contain information about properties of the model and allows us to compare our result against certain trivial limiting cases.
For example, if $b=0$, then $\rho(\lambda) = D_{\rm H}\delta(\lambda)$ and $a_1=0$ in accordance with Eq.~\eqref{eq:a0a1}.
At the same time, if $b=1$ ($\bar b=0$), then $\rho(\lambda) = D_{\rm H}\delta(\lambda-1)$ and $a_1=1$, again in accordance with Eq.~\eqref{eq:a0a1}.
Other instructive limits to consider are $\gamma\to 0 $ and $\gamma\to\infty$.
In the former limit, matrix $U$ defined in Eq.~\eqref{eq:cayley} becomes $U=\openone$, while in the latter $U=-\openone$.
In both cases $\Lambda_P = P$ and hence $\rho(\lambda) = D_{\rm H}\left[ b \delta(\lambda-1) +\bar b \delta(\lambda)\right]$, which yields $a_1=b$, again in accordance with Eq.~\eqref{eq:a0a1}.
Final important limiting case corresponds to $\gamma=2$.
As we mentioned earlier [see discussion below Eq.~\eqref{eq:cayley}], in this case the unitary ensemble we consider mimics the circular unitary ensemble (CUE) and $a_1=b^2$ in accordance with Eq.~\eqref{eq:a0a1}.
Further comparison between our calculations for $\gamma=2$ and the CUE case are given in \ref{sec:App_Eq_61}.

\subsection{Analysis near $\lambda=1$ and the phase transition}

The limit of $\lambda\to1$ corresponds to $f_-\to+\infty$, $f_+\to 0^+$ and $f_+f_-=1$.
In this case, system of equations \eqref{eq:sc_vpm} can be solved by the following ansatz:
\begin{equation}
	v_-|_{\lambda\to1^-}=f_-x+\sqrt{f_-^2x^2+f_-y},
	\label{eq:vm_ansat}
\end{equation}
where $x,y$ are unknown functions of $\gamma,b$.
Note that the expansion of Eq.~\eqref{eq:vm_ansat} in the limit $f_-\to+\infty$ depends on the sign of $x$.

Substituting ansatz \eqref{eq:vm_ansat} into Eqs.~\eqref{eq:sc_vpm} and expanding the corresponding fifth-order equation on $v_-$ up to the leading order in $1/f_-$ allows us to obtain expressions for $x$ and $y$:
\begin{subequations}
\begin{align}
	&x=\frac{1}{4}\left[-(\gamma\bar{b}+1)+\sqrt{(\gamma\bar b-1)^2+4\gamma b}\right],\label{eq:xi} \\
	&y=-x\frac{2 b+\bar b\left[\gamma b-1+\sqrt{4\gamma\bar b+(\gamma b-1)^2} \right]}{\bar b- b}.\label{eq:eta}
\end{align}
\end{subequations}
Note that $x=0$ when $b=\bar b =1/2$.
As a next step, we use expression \eqref{eq:vm_ansat} for $v_-$ with $x,y$ given by Eqs.~\eqref{eq:xi},\eqref{eq:eta} in Eq.~\eqref{eq:F}.
In the general case, the expression for $\mathcal{F}$ after such substitution is cumbersome.
For this reason, we restrict ourselves to computing it in the vicinity of $b=\bar b=1/2$.
In this case, we obtain
\begin{equation}
	\mathcal{F}=\frac{4\gamma-\sqrt{2\gamma(\gamma+2)^2\lambda-2\gamma[32\gamma b\bar b + (\gamma-2)^2]}}{2\gamma(\lambda-1)}.
    \label{eq:F_near_trans}
\end{equation}
Substituting Eq.~\eqref{eq:F_near_trans} into Eq.~\eqref{eq:Tr_G_pm} we obtain
\begin{align}
	\Tr_{\rm H}[G(\lambda)]&= D_{\rm H}\left\{1+\frac{b-\bar b}{2(\lambda-1)}+\frac{\sqrt{(\gamma+2)^2\lambda-[32\gamma b\bar b + (\gamma-2)^2]}}{4\sqrt{2\gamma}(\lambda-1)}\right\}.\label{eq:Tr_G_lambda_11}
\end{align}
Note that in the limiting case $\gamma=2$ this expression transforms into:
\begin{align}
	\Tr_{\rm H}[G(\lambda)]&= D_{\rm H}\left\{1+\frac{b-\bar b}{2(\lambda-1)}+\frac{\sqrt{\lambda-4 b\bar b}}{2(\lambda-1)}\right\}.
\end{align}
This limit is consistent with the result \eqref{eq:rho_lambda_math} for the Haar-random case derived using free probability theory techniques in Ref.~\cite{Collins2005}.

Next, we substitute expression \eqref{eq:Tr_G_lambda_11} into Eq.~\eqref{eq:dos} to find the density of states for the case $|\lambda-1|\ll1 $ and general values of $\gamma$:
\begin{align}
    \frac{\rho(\lambda)}{D_H}&=(b-\bar b)\theta(b-\bar b)\delta(\lambda-1)+\frac{\gamma+2}{4\pi\sqrt{2\gamma}}\frac{\sqrt{1-\lambda^{\ast}-\lambda}}{1-\lambda}\theta(\lambda)\theta\left[1-\lambda^{\ast}-\lambda\right],\label{eq:rho_lambda_1}
\end{align}
where $\theta(x)$ is the Heaviside step function and
\begin{equation}
    \lambda^{\ast} = \frac{8\gamma}{(\gamma+2)^2}\left(1-4 b\bar b\right).
    \label{eq:d_ast_1}
\end{equation}
Here $\lambda^{\ast}$ is nothing but the gap in the spectrum defined in Eq.~\eqref{eq:w_d} and it reveals itself as the decay rate of the probability $w(d)$.
Moreover, we find that the density of $\lambda=1$ eigenvalues for $b>\bar b$ equals $P(\lambda=1)=b-\bar b$, which determines the steady-state probability $w_{\infty}$ through Eq.~\eqref{eq:w_inf_P_1}.
The probability $w_{\infty}$ is thus independent of the value of $\gamma$.
In addition, expression \eqref{eq:rho_lambda_1} gives the critical exponent $\eta=1/2$ in Eq.~\eqref{eq:w_d_0}.
Thus, $\gamma$ affects the coefficients but does not have an impact on the critical behavior.

\begin{figure}[t]
    \centering
    \includegraphics[width = 0.98\columnwidth]{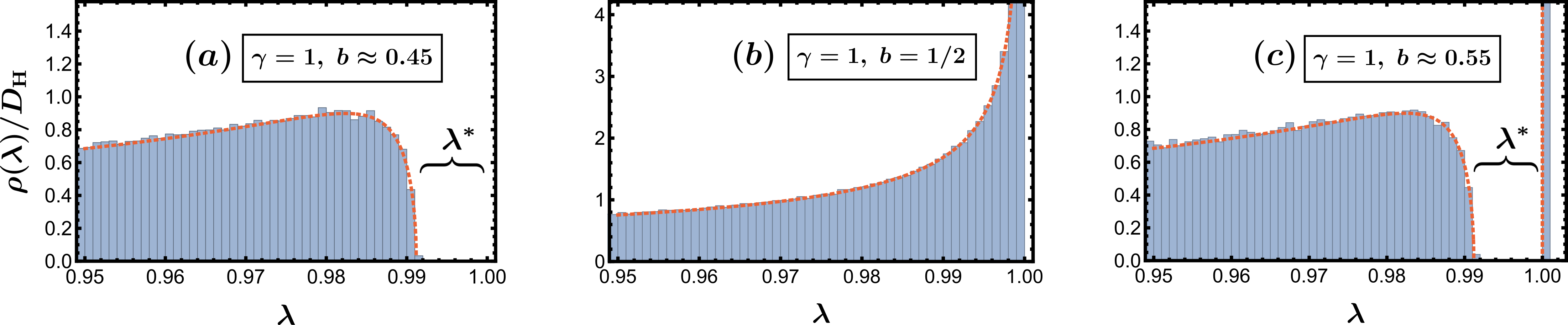}
    \caption{Histograms representing the density of states of a numerically generated random matrix $\Lambda_P$ for $\gamma=1$ and various values of $b$ plotted near $\lambda=1$. The dimension of the Hilbert space is $D_{\rm H}=2048$. Averaging is performed over 400, 800, and 400 random matrix realizations in panels (a),(b), and (c), respectively. The red curve shows the analytically derived expression \eqref{eq:rho_lambda_1} for the density of states near $\lambda=1$ in the vicinity of the critical point $b=1/2$.
    }
    \label{fig:W_eval_hist_eval1_1}
\end{figure}

Figure~\ref{fig:W_eval_hist_eval1_1} depicts the density of states calculated numerically for random matrices as well as the expression \eqref{eq:rho_lambda_1}, showcasing a good agreement between our analytical results and the numerics.

Equations~\eqref{eq:rho_lambda_1}-\eqref{eq:d_ast_1} form a complete solution for the density of states in the vicinity of the phase transition, $b=\bar b$, and fully characterize the phase transition of Eq.~\eqref{eq:w_d}.
This phase transition has been obtained previously for the Haar-random case~\cite{Collins2005}.
Our results generalize these results to the case of finite scrambling $\gamma\neq 2$.

\subsection{Analysis near $\lambda=0$ and one more phase transition}

\begin{figure}[t]
    \centering
    \includegraphics[width = 0.4\columnwidth]{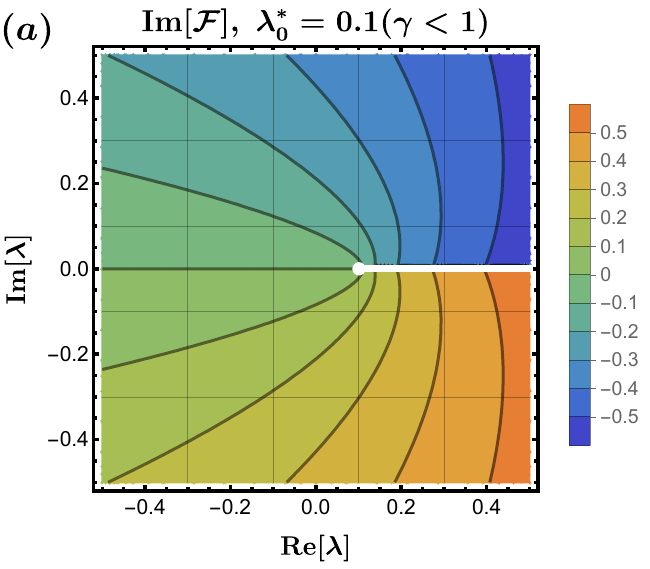}\qquad\qquad
    \includegraphics[width = 0.4\columnwidth]{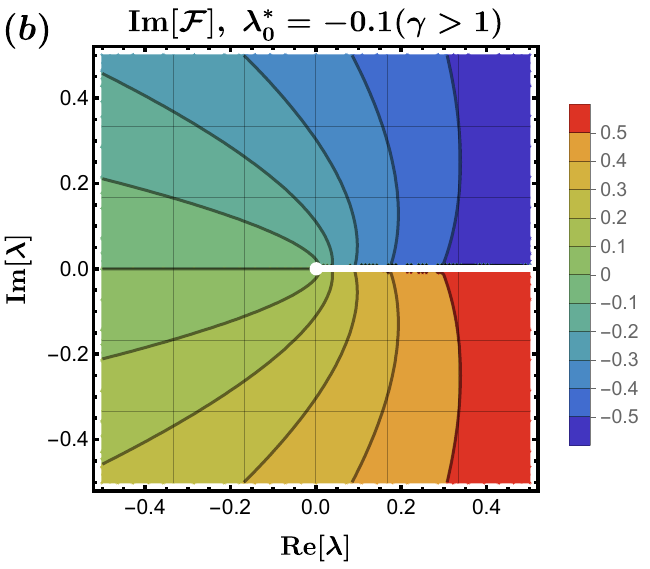}
    \caption{Function \text{Im}$[\mathcal{F}]$ given by Eq.~\eqref{eq:F_near_lambda0} plotted as a function of $\lambda$ on a complex plain for (a) $\lambda_0^{\ast}=0.1(\gamma<1)$ and (b) $ \lambda_0^{\ast}=-0.1(\gamma>1)$. In both cases the function is analytical everywhere except for the corresponding branch cuts depicted as white solid lines.}
    \label{fig:F_complex_plain}
\end{figure}

\begin{figure}[t]
    \centering
    \includegraphics[width = 1\columnwidth]{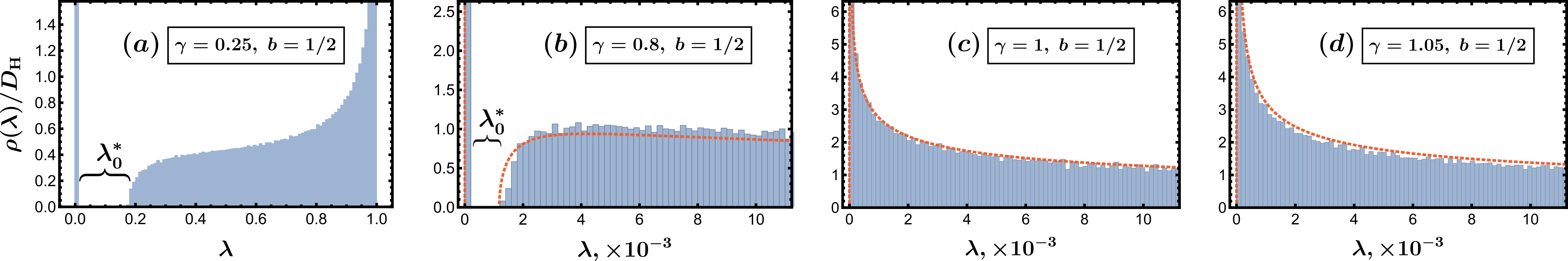}
    \caption{Histograms representing the density of states of a numerically generated random matrix $\Lambda_P$ for $b=1/2$ and various values of $\gamma$. The dimension of the Hilbert space is $D_{\rm H}=2048$. Averaging is performed over 40, 1600, 800, and 800 random matrix realizations in panels (a),(b), (c), and (d), respectively. The red curve shows the analytically derived expression [see Eqs.~\eqref{eq:dos},\eqref{eq:Tr_G_pm} and \eqref{eq:F_near_lambda0}-\eqref{eq:lambda_ast}] for the density of states near $\lambda=0$ in the vicinity of the critical point $\gamma=1$. In panels (b), (c) and (d) the $x$-axis scale is zoomed in near $\lambda=0$ to resolve features in the spectrum close to the transition and to have a better agreement between numerical and analytical results.
    }
    \label{fig:W_eval_hist_eval0_1}
\end{figure}

In the limit $\lambda\to 0$ it follows that $f_{\pm}\to -1$.
For general $b$, the solution of Eqs.~\eqref{eq:sc_vpm} and the expression for $\mathcal{F}$ in this limit are cumbersome and for simplicity we focus on the case when $b=\bar b=1/2$.
Then, solving Eqs.~\eqref{eq:sc_vpm} for $v_{\pm}$ and plugging the result into expression \eqref{eq:F} in the limit $\gamma\to 1$, $\lambda\to0$ we obtain
\begin{align}
    &\mathcal{F} = \frac{1}{2^{2/3}}\left\{\left|\lambda_0^{\ast}\right|^{1/3}\text{sign}\left(\lambda_0^{\ast}\right)+\left[\left(\sqrt{\lambda_0^{\ast}-\lambda}+\sqrt{-\lambda}\right)^{2}\right]^{\frac{1}{3}}+\left[\left(\sqrt{\lambda_0^{\ast}-\lambda}-\sqrt{-\lambda}\right)^{2}\right]^{\frac{1}{3}} \right\},\label{eq:F_near_lambda0}
\end{align}
where we have introduced 
\begin{equation}
    \lambda_0^{\ast} = \frac{4}{27}(1-\gamma)^3.
    \label{eq:lambda_ast}
\end{equation}
To compute the density of states \eqref{eq:dos},\eqref{eq:Tr_G_pm}, we need to study properties of $\mathcal{F}$ as a function of $\lambda$ on a complex plain.
To illustrate these properties, we depict $\text{Im}[\mathcal{F}](\lambda)$ in Fig.~\ref{fig:F_complex_plain}.
When $ \lambda_0^{\ast}>0 (\gamma<1)$, square roots $\sqrt{-\lambda}$ in Eq.~\eqref{eq:F_near_lambda0} cancel out and function $\mathcal{F}(\lambda)$ is analytical on a complex plain except for a line $\text{Re}(\lambda)> \lambda_0^{\ast}$, $\text{Im}(\lambda)=0$, where it has a branch cut.
The corresponding case is illustrated in Fig.~\ref{fig:F_complex_plain}(a), where the branch cut is depicted as a white line.

On the other hand, when $ \lambda_0^{\ast}<0 (\gamma>1)$, square roots $\sqrt{ \lambda_0^{\ast}-\lambda}$ in Eq.~\eqref{eq:F_near_lambda0} cancel out, and the branch cut is located on a line $\text{Re}(\lambda)>0$, $\text{Im}(\lambda)=0$, see Fig.~\ref{fig:F_complex_plain}(b).

The above analysis indicates that for $\gamma<1$ the density of states $\rho(\lambda)$ has a gap near $\lambda=0$ that is equal to $\lambda_0^{\ast}$.
In contrast, for $\gamma>1$ the spectrum near $\lambda=0$ is gapless. 
This suggests the presence of the phase transition at $\gamma=1$ between a gapped and a gapless phase.
Figure~\ref{fig:W_eval_hist_eval0_1} depicts the density of states calculated using numerically generated random matrices as well as analytical expressions \eqref{eq:dos},\eqref{eq:Tr_G_pm} and \eqref{eq:F_near_lambda0}-\eqref{eq:lambda_ast}, demonstrating the presence of the phase transition in the spectrum as well as a good agreement between the random-matrix numerical results and our analytical analysis.

Within the gapped phase ($\gamma<1$), expanding Eq.~\eqref{eq:F_near_lambda0} near $\lambda=\lambda_0^{\ast}$ we find the following expression for the density of states:
\begin{align}
    \rho(\lambda)&=D_{\rm H}\left\{\bar b\delta(\lambda)+\frac{9}{4\pi(1-\gamma)^{5/2}}\theta\left[\lambda-\lambda_0^{\ast}\right]\sqrt{\lambda-\lambda_0^{\ast}}\right\},
    \label{eq:rho_lambda_0_gamma<1}
\end{align}
where $\theta(x)$ is the Heaviside step function.
Here the first term is due to eigenvalues at $\lambda=0$ of the projector $P$, while the second term describes the edge of the continuum spectrum.
On the other hand, within the gapless phase ($\gamma>1$), expanding near $\lambda=0$ we find
\begin{equation}
    \rho(\lambda)=D_{\rm H}\left[\bar b\delta(\lambda)+\frac{1}{\pi}\theta(\lambda)\sqrt{\frac{\gamma-1}{\lambda}}\right],
    \label{eq:rho_lambda_0_gamma>1}
\end{equation}
while at the critical point, $\gamma=1$, we obtain
\begin{equation}
    \rho(\lambda)=D_{\rm H}\left[\bar b\delta(\lambda)+\frac{\sqrt{3}}{2\pi}\theta(\lambda)\frac{1}{\lambda^{1/3}}\right].
    \label{eq:rho_lambda_0_gamma=1}
\end{equation}

So far our analysis of the spectrum near $\lambda=0$ has focused on the case $b=\bar b=1/2$.
For general values of $b$ the corresponding analytical expressions are cumbersome and we elect not to present them here.
However, we do observe that the gapped-to-gapless phase transition similar to the one depicted in Fig.~\ref{fig:W_eval_hist_eval0_1} is still present in the general case.
The critical value $\gamma_c(b)$ is given by
\begin{equation}
	\gamma_c(b)=\frac{1/2-\sqrt{b(1-b)}}{\left(b-1/2\right)^2},
	\label{eq:gamma_c}
\end{equation}
while the functional dependence of $\rho(\lambda)$ near the critical point remains the same as in Eqs.~\eqref{eq:rho_lambda_0_gamma<1}-\eqref{eq:rho_lambda_0_gamma=1}.

Naively, one could think that the transition near $\lambda=0$ analyzed in this subsection and depicted in Fig.~\ref{fig:W_eval_hist_eval0_1} cannot be experimentally observed.
Indeed, only eigenvalues close to $\lambda=1$ can have an effect on the quantum dynamics described by $\Lambda_P^d$, which is reflected in the probabilities \eqref{eq:w_d} to get the same measurement outcome.
However, the transition \textit{can} be observed if one considers the dynamics generated by an operator 
\begin{equation}
    \Lambda_{01}=PU^{\dag}\bar PUP=\openone-\Lambda_P,
\end{equation}
where $\bar P=\openone-P$ is the projection operator corresponding to the measurement outcome (which we call ``1") different from the measurement outcome of the projector $P$ (which we call ``0").
The spectrum of the operator $\Lambda_{01}$ is then equal to the spectrum of $\Lambda_{P}$ mirror-reflected around $\lambda=1/2$.
The transition near eigenvalue $\lambda=0$ in the spectrum of $\Lambda_{P}$ (depicted in Fig.~\ref{fig:W_eval_hist_eval0_1}) thus becomes a transition near eigenvalue $\lambda=1$ in the spectrum of $\Lambda_{01}$, and therefore can be observed similarly to the transition of Eq.~\eqref{eq:w_d}.
Specifically, if one considers the probability of a quantum trajectory ``0101\dots 010" given by
\begin{equation}
    w_{01}(d)=w_{\underbrace{\scriptstyle0101\dots010}_\text{$(d+1)$}}=\Tr[\Lambda_{01}^{d}\rho_0],
\end{equation}
its asymptotic behavior in the limit $d\gg1$ reads
\begin{subequations}\label{eq:w_1_d}
\begin{align}
    w_{01}(d) &\sim \exp(-\lambda_0^{\ast}d),\quad \text{for ${\gamma<\gamma_c}$},   \\
    w_{01}(d) &\sim \frac{1}{d^{\eta_1}},\quad \text{for ${\gamma=\gamma_c}$},   \\
    w_{01}(d) &\sim \frac{1}{d^{\eta_2}},\quad \text{for ${\gamma>\gamma_c}$},
\end{align}
\end{subequations}
where $\lambda_0^{\ast}$ is the gap in the spectrum [given by Eq.~\eqref{eq:lambda_ast} for $b=1/2$] that determines the decay rate of the probability, $\gamma_c$ is given by Eq.~\eqref{eq:gamma_c} and the critical exponents are $\eta_1=1/3$, $\eta_2=1/2$.
The phase transition can thus be identified by observing different scalings of the probability \eqref{eq:w_1_d}.

In contrast to the transition in $P(\lambda=1)$, we did not observe the transition in $P(\lambda=0)$ numerically in one- and two-dimensional grids as the randomization of the wave functions cannot be described there by a simple parameter $\gamma$.
We have numerical evidence that such transition can be observed in different operators such as 
\begin{equation}
 \Lambda_{PP'}=PU^\dagger P'U,
\end{equation}  
in which projectors $P$ and $P'$ act on different qubit clusters.

\section{\label{sec:Conclusion}Conclusion}

In this paper we introduced an alternative measurement-induced transition that can in principle be observed in an experiment without requiring an exponential overhead in resources.
Our proposal is based on the spectral properties of an operator $\Lambda_P=PU^{\dag}PUP$, where we characterize the projector $P$ by a parameter $b$ such that $\Tr[P]=bD_{\rm H}$.
Studying $\Lambda_P$ analytically for the case when $U$ is a random unitary matrix and numerically for the case when $U$ is constructed by a repeated application of local quantum gates in a one- and two-dimensional qubit grid, we discovered that the spectrum of $\Lambda_P$ exhibits a phase transition at a critical value of $b=b_c$.
Specifically, for $b>b_c$ the spectrum of $\Lambda_P$ possesses a nonzero density of eigenvalues at $\lambda=1$, whereas for $b<b_c$ the eigenvalues at $\lambda=1$ are absent.
When $U$ is a random unitary matrix, we analytically calculated $b_c=0.5$.
While the analytical results for a Haar-random matrix $U$ have been previously reported \cite{Collins2005}, in this work we generalize them to the case when the degree of scrambling by $U$ is an adjustable parameter [defined by $\gamma$ in Eqs.~\eqref{eq:GUE}-\eqref{eq:cayley}].
Strikingly, we found that $b_c$ is independent of $\gamma$.
In addition, our numerics in one and two dimensions show consistency with the analytical result, giving $b_c=0.5$ in the former case and $b_c\approx 0.5-0.6$ in the latter.
Unfortunately, the available system sizes in the two-dimensional case do not allow us to extract the precise location of the critical point.

Additionally, we discovered that for the random-matrix model the spectrum of $\Lambda_P$ is separated from $\lambda=1$ by a finite gap when $b>b_c$ and $b<b_c$, whereas at the transition, $b=b_c$, this gap closes.
We further derived analytical expressions \eqref{eq:rho_lambda_1},\eqref{eq:d_ast_1} for the density of states and the gap near the transition in the vicinity of $\lambda=1$, which coincide with the random-matrix numerics.
In contrast to the location of the transition $b_c$, the value of the gap does depend on the scrambling parameter $\gamma$.
For the one-dimensional model we found a strong suppression of the spectral density near $\lambda=1$ when $b>b_c$ and $b<b_c$.
The suppression grows stronger with the number of layers in the quantum circuit comprising the unitary $U$.
When the number of layers is equal to the system size, the one-dimensional spectrum closely resembles the spectrum of the random-matrix model.
In contrast, in the two-dimensional case the spectral density near $\lambda=1$ remains finite (albeit small) for all values of $b$, which could be due to the finite size effects.
We leave the analysis of this phenomenon for future studies.

The transition in the spectrum of $\Lambda_P$ can be observed experimentally because it translates into the distinct asymptotic behavior for a probability $w$ of a quantum trajectory produced by a repeated application of $U$ and $U^{\dag}$ interspersed by the same projective measurements where only one measurement outcome (corresponding to the projector $P$) is accepted.
This behavior of the probability is described by Eq.~\eqref{eq:w_d}.
For $b<b_c$, the absence of the eigenvalues at $\lambda=1$ in the spectrum of  $\Lambda_P$ gives way to the exponential decay to zero of the probability $w$ with the number of measurements.
The gap in the spectrum then determines the decay rate.
Such exponential decay of the trajectory probability is also a feature of generic quantum circuits subjected to repeated measurements.
On the other hand, for $b>b_c$ the probability $w$ decays to a constant value $w_{\infty}$ due to the presence of the eigenvalues at $\lambda=1$.
The value of $w_{\infty}$ in this case is characterized by the number of $\lambda=1$ eigenvalues.
Meanwhile, at the critical point, $b=b_c$, the probability decays as a power-law with a critical exponent $\eta=1/2$.
These predictions can be tested experimentally by the direct measurement of the outcome probabilities for the quantum evolution on a two dimensional grid. The crucial requirement for the experimental observation is the ability to produce the forward evolution and its perfect time reversal. This is a non-trivial task that became only recently possible \cite{Mi2021}.

\section*{\label{sec:D_of_I}Declaration of competing interest}

The authors declare that they have no known competing financial interests or personal relationships that could have appeared to influence the work reported in this paper.

\section*{\label{sec:acknowledgments}Acknowledgments}

We thank Rosario Fazio, Michael Freedman, Gerald Fux, Alexey Milekhin, Mark Srednicki, Yaodong Li, Thomas Iadecola and Kostyantyn Kechedzhi for useful discussions. A.K. was partially supported by the U.S. Department of Energy, Office of Science, Basic Energy Sciences, Materials Science and Engineering Division, including the grant of computer time at the National Energy Research Scientific Computing Center (NERSC) in Berkeley, California. This part of research was performed at the Ames National Laboratory, which is operated for the U.S. DOE by Iowa State University under Contract No. DE-AC02-07CH11358.

\appendix
\section{\label{sec:App_Eq_27}Derivation of Equation~\eqref{eq:series1} of the main text}

Matrix $\hat{\mathcal{M}}(\xi)$ of Eq.~\eqref{eq:M_lambda_def} can be written as
\begin{equation}
	\hat{\mathcal{M}}(\xi) =(\hat 1+\sqrt{\xi}\hat{\mathcal{T}}\hat{\mathcal{L}}^{-1})\hat{\mathcal{L}},
\end{equation}
and thus its inverse reads
\begin{equation}
	\hat{\mathcal{M}}^{-1}(\xi)=\hat{\mathcal{L}}^{-1}(\hat 1+\sqrt{\xi}\hat{\mathcal{T}}\hat{\mathcal{L}}^{-1})^{-1}=\hat{\mathcal{L}}^{-1}+\hat{\mathcal{L}}^{-1}(-\sqrt{\xi}\hat{\mathcal{T}}+\xi\hat{\mathcal{T}}\hat{\mathcal{L}}^{-1}\hat{\mathcal{T}})\sum_{k=0}^{\infty}\xi^{k}(\hat{\mathcal{L}}^{-1}\hat{\mathcal{T}}\hat{\mathcal{L}}^{-1}\hat{\mathcal{T}})^k \hat{\mathcal{L}}^{-1},
	\label{eq:M_inter}
\end{equation}
where
\begin{equation}
	\hat{\mathcal{L}}^{-1}=
	\begin{pmatrix}
		\check 0 & (\check L^{\dagger})^{-1} \\
		\check L^{-1} & \check 0
	\end{pmatrix}_{\rm K}. \label{eq:matcal_L_inverse}
\end{equation}
Using Eq.~\eqref{eq:matcal_L_inverse} and the definition of the matrix $\hat{\mathcal{T}}$ given by the second part of Eq.~\eqref{eq:M_lambda_def}, it is straightforward to show that
\begin{align}
	\hat{\mathcal{L}}^{-1}\hat{\mathcal{T}}\hat{\mathcal{L}}^{-1}\hat{\mathcal{T}}=
	\begin{pmatrix}
		(\check L^{\dagger})^{-1}\check t_2\check L^{-1}\check t_1  & \check 0 \\
		\check 0 & \check L^{-1}\check t_1(\check L^{\dagger})^{-1}\check t_2
	\end{pmatrix}_{\rm K}
\end{align}
and
\begin{align}
	&\Tr_{\rm K,\rm A,\rm H}\left[\hat{\mathcal{T}}\hat{\mathcal{L}}^{-1}\right]=0 \label{eq:tr_tl_1}, \\ 
	&\Tr_{\rm K,\rm A,\rm H}\left[\left(\hat{\mathcal{L}}^{-1}\hat{\mathcal{T}}\hat{\mathcal{L}}^{-1}\hat{\mathcal{T}}\right)^k\right]=2\Tr_{\rm A,\rm H}\left[\left((\check L^{\dagger})^{-1}\check t_2\check L^{-1}\check t_1\right)^k\right], \\
	&\Tr_{\rm K,\rm A,\rm H}\left[ \hat{\mathcal{L}}^{-1}\hat{\mathcal{T}}\left(\hat{\mathcal{L}}^{-1}\hat{\mathcal{T}}\hat{\mathcal{L}}^{-1}\hat{\mathcal{T}}\right)^k\right]=0. \label{eq:tr_tl_4}
\end{align}
Next, we consider a function $\sqrt{\xi}\Tr_{\rm K,\rm A,\rm H}\left[\hat{\mathcal{T}}\hat{\mathcal{M}}^{-1}\right]$, which with the help of Eqs.~\eqref{eq:tr_tl_1}-\eqref{eq:tr_tl_4} can be written as
\begin{align}
	\sqrt{\xi}\Tr_{\rm K,\rm A,\rm H}\left[\hat{\mathcal{T}}\hat{\mathcal{M}}^{-1}\right]&=\sqrt{\xi}\Tr_{\rm K,\rm A,\rm H}\left[\hat{\mathcal{T}}\hat{\mathcal{L}}^{-1}\hat{\mathcal{T}}(-\sqrt{\xi}\hat 1+\xi\hat{\mathcal{L}}^{-1}\hat{\mathcal{T}})\sum_{k=0}^{\infty}\xi^{k}(\hat{\mathcal{L}}^{-1}\hat{\mathcal{T}}\hat{\mathcal{L}}^{-1}\hat{\mathcal{T}})^k \hat{\mathcal{L}}^{-1}\right]\nonumber\\
	&=\Tr_{\rm K,\rm A,\rm H}\left[(-\hat 1+\sqrt{\xi}\hat{\mathcal{L}}^{-1}\hat{\mathcal{T}})\sum_{k=0}^{\infty}\xi^{k+1}(\hat{\mathcal{L}}^{-1}\hat{\mathcal{T}}\hat{\mathcal{L}}^{-1}\hat{\mathcal{T}})^k \hat{\mathcal{L}}^{-1}\hat{\mathcal{T}}\hat{\mathcal{L}}^{-1}\hat{\mathcal{T}}\right]\nonumber\\
	&=\Tr_{\rm K,\rm A,\rm H}\left[(-\hat 1+\sqrt{\xi}\hat{\mathcal{L}}^{-1}\hat{\mathcal{T}})\sum_{k=1}^{\infty}\xi^{k}(\hat{\mathcal{L}}^{-1}\hat{\mathcal{T}}\hat{\mathcal{L}}^{-1}\hat{\mathcal{T}})^k \right]\nonumber\\
	&=-2\sum_{k=1}^{\infty}\xi^{k}\Tr_{\rm A,\rm H}\left[\left((\check L^{\dagger})^{-1}\check t_2\check L^{-1}\check t_1\right)^k \right].\label{eq:inter1}
\end{align}
This is nothing but Eq.~\eqref{eq:series1} of the main text.

\section{\label{sec:App_Eq_42}Derivation of Equation~\eqref{eq:g0m1} of the main text}

From Eq.~\eqref{eq:M_inter} it follows that
\begin{equation}
    \hat{\mathcal{G}}_0(\xi)\equiv\hat{\mathcal{M}}^{-1}_0(\xi)=\hat{\mathcal{L}}^{-1}_0+\hat{\mathcal{L}}^{-1}_0(-\sqrt{\xi}\hat{\mathcal{T}}+\xi\hat{\mathcal{T}}\hat{\mathcal{L}}^{-1}_0\hat{\mathcal{T}})\sum_{k=0}^{\infty}\xi^{k}(\hat{\mathcal{L}}^{-1}_0\hat{\mathcal{T}}\hat{\mathcal{L}}^{-1}_0\hat{\mathcal{T}})^k \hat{\mathcal{L}}^{-1}_0,
	\label{eq:G0_def}
\end{equation}
where $\hat{\mathcal{L}}^{-1}_0=\hat{\mathcal{L}}^{-1}|_{H=0}$.
Using explicit forms of matrices $\hat{\mathcal{T}}$ and $\hat{\mathcal{L}}^{-1}_0$, see Eqs.~\eqref{eq:M_lambda_def},\eqref{eq:t12} and \eqref{eq:matcal_L_inverse}, we obtain the following identity: 
\begin{equation}
	\hat{\mathcal{T}}\hat{\mathcal{L}}_0^{-1}\hat{\mathcal{T}}\hat{\mathcal{L}}_0^{-1}\hat{\mathcal{T}}=\hat{\mathcal{T}},
	\label{eq:TLTLT}
\end{equation}
using which we simplify expression \eqref{eq:G0_def} for $\hat{\mathcal{G}}_0$ into
\begin{equation}
	\hat{\mathcal{G}}_0=\hat{\mathcal{L}}^{-1}_0+\hat{\mathcal{L}}^{-1}_0(-\sqrt{\xi}\hat{\mathcal{T}}+\xi\hat{\mathcal{T}}\hat{\mathcal{L}}^{-1}_0\hat{\mathcal{T}})\sum_{k=0}^{\infty}\xi^{k} \hat{\mathcal{L}}^{-1}_0=\hat{\mathcal{L}}^{-1}_0+\hat{\mathcal{L}}^{-1}_0\left(-\frac{\sqrt{\xi}\hat{\mathcal{T}}}{1-\xi}+\frac{\xi\hat{\mathcal{T}}\hat{\mathcal{L}}_0^{-1}\hat{\mathcal{T}}}{1-\xi}\right)\hat{\mathcal{L}}^{-1}_0.
	\label{eq:G0_inter}
\end{equation}
Next, using the projector $\hat\Pi$ onto the space $\mathbb H_{\rm I}$, see Eq.~\eqref{eq:Pi}, we obtain the following identities:
\begin{align}
	&\hat\Pi\hat{\mathcal{L}}_0^{-1}\hat\Pi=
	\begin{pmatrix}
		0 & 0 & 0 & 0 \\
		0 & 0 & -i & 0 \\
		0 & i & 0 & 0 \\
		0 & 0 & 0 & 0 \\
	\end{pmatrix}_{\rm K,A}, \label{eq:Pi_L0m1_Pi}\\
	&\hat\Pi\hat{\mathcal{L}}_0^{-1}\hat{\mathcal{T}}\hat{\mathcal{L}}_0^{-1}\hat\Pi=2P
	\begin{pmatrix}
		0 & 0 & 0 & 0 \\
		0 & 1 & 0 & 0 \\
		0 & 0 & 1 & 0 \\
		0 & 0 & 0 & 0 \\
	\end{pmatrix}_{\rm K,A}, \\
	&\hat\Pi\hat{\mathcal{L}}_0^{-1}\hat{\mathcal{T}}\hat{\mathcal{L}}_0^{-1}\hat{\mathcal{T}}\hat{\mathcal{L}}_0^{-1}\hat\Pi=2P
	\begin{pmatrix}
		0 & 0 & 0 & 0 \\
		0 & 0 & -i & 0 \\
		0 & i & 0 & 0 \\
		0 & 0 & 0 & 0 \\
	\end{pmatrix}_{\rm K,A}. \label{eq:Pi_L0m1TL0m1TL0m1_Pi}
\end{align}
We utilize identities \eqref{eq:Pi_L0m1_Pi}-\eqref{eq:Pi_L0m1TL0m1TL0m1_Pi} to project Eq.~\eqref{eq:G0_inter} onto the space $\mathbb H_{\rm I}$, which yields
\begin{equation}
	g_0=P\left(f_2\tau_2 +f_0\tau_0\right)+\bar P\tau_2,
	\label{eq:G0_final}
\end{equation}
where $\bar P = \openone-P$, $\{\tau_i\}_{i=0}^4$ are the standard Pauli matrices in the space $\mathbb H_{\rm I}$, and
\begin{align}
	f_2=\frac{1+\xi}{1-\xi},\quad f_0=-\frac{2\sqrt{\xi}}{1-\xi}.
\end{align}
Expression \eqref{eq:g0m1} of the main text is then obtained by inverting $g_0$ of Eq.~\eqref{eq:G0_final}:
\begin{equation}
	g_0^{-1}=P(f_2\tau_2 -f_0 \tau_0)+\bar P\tau_2.
\end{equation}

\section{\label{sec:App_Eq_55_56}Derivation of Equations~\eqref{eq:tr_g0_t},\eqref{eq:pi_g0_t_g0_pi} of the main text}

We start by considering expression~\eqref{eq:G0_inter} for $\hat{\mathcal{G}}_0$.
Using Eq.~\eqref{eq:tr_tl_1} with $\hat{\mathcal{L}}_0^{-1}$, Eq.~\eqref{eq:TLTLT} and the fact that 
\begin{equation}
	\hat{\mathcal{L}}_0^{-1}\hat{\mathcal{T}}\hat{\mathcal{L}}_0^{-1}\hat{\mathcal{T}}=P
	\begin{pmatrix}
		1 & 0 & 0 & 0 \\
		-2i & 0 & 0 & 0 \\
		0 & 0 & 1 & -1 \\
		0 & 0 & 0 & 0 \\
	\end{pmatrix}_{\rm K,A},
\end{equation}
in Eq.~\eqref{eq:G0_inter}, we obtain Eq.~\eqref{eq:tr_g0_t} of the main text:
\begin{equation}
	\Tr_{\rm K,\rm A,\rm H}\left[\hat{\mathcal{T}}\hat{\mathcal{G}}_0\right]= f_0 b D_{\rm H},
\end{equation}
where $f_0$ has been introduced earlier, see Eq.~\eqref{eq:f2f0}.

Next, to derive Eq.~\eqref{eq:pi_g0_t_g0_pi} of the main text, we recall Eq.~\eqref{eq:M_lambda_def} and note that
\begin{equation}
	\hat{\mathcal{T}}=\frac{\partial\hat{\mathcal{M}}_0}{\partial\sqrt\xi}=\frac{\partial\hat{\mathcal{G}}_0^{-1}}{\partial\sqrt\xi}=-\hat{\mathcal{G}}_0^{-1}\frac{\partial\hat{\mathcal{G}}_0}{\partial\sqrt\xi}\hat{\mathcal{G}}_0^{-1}.
\end{equation}
Hence,
\begin{equation}
	\eta=-\frac{\partial g_0}{\partial\sqrt\xi}=\frac{2P}{1-\xi}\left(f_0\tau_2 +f_2\tau_0\right),
\end{equation}
which is nothing but Eq.~\eqref{eq:pi_g0_t_g0_pi} of the main text.

\section{\label{sec:App_Eq_61} Calculation of the first three moments of the random-matrix distribution and comparison with the circular unitary ensemble (CUE) case}

Setting $\lambda\to\infty$, $f_{\pm}=1$ in Eqs.~\eqref{eq:sc_vpm} gives
\begin{align}
	\begin{cases}
		v_+=\dfrac{\gamma}{v_-+1}, \\
		v_-=\dfrac{\gamma}{v_++1}.
	\end{cases}
 \label{eq:vp_vm_laminf}
\end{align}
Solution of the system of equations~\eqref{eq:vp_vm_laminf} reads
\begin{equation}
	v_+=v_-\equiv v=\frac{1}{2}\left(-1+\sqrt{1+4\gamma}\right).
	\label{eq:v_pm_sol_lam0}
\end{equation}
Here we pick a positive sign to match the moments of the random-matrix distribution in certain known limits (see the main text and the discussion below).
Expression \eqref{eq:F} for $\mathcal F$ in the case of $\lambda\to\infty$ can be rewritten as
\begin{equation}
	\mathcal F=(v+1)^4-\gamma(v+1)^2=(v+1)^3,
\end{equation}
and thus using Eq.~\eqref{eq:Tr_G_pm} we find
\begin{equation}
	\Tr_{\rm H}[G(\lambda)]= D_{\rm H}\left[\frac{a_0}{\lambda}+\frac{a_1}{\lambda^2} \right]
\end{equation}
with the moments $a_0,a_1$ given by Eq.~\eqref{eq:a0a1}.

At the same time, for the case when the matrix $U$ entering the expression~\eqref{eq:Lambda_P} for $\Lambda_P$ is sampled from the CUE, we perform random-Haar averaging and find $a_1=b^2$, consistent with the result \eqref{eq:a0a1} for our model when $\gamma=2$.
We further perform the expansion of $\Tr_{\rm H}[G(\lambda)]$ in the powers of $1/\lambda$ to the third and the fourth orders, which for $\gamma=2$ yields
\begin{equation}
    a_2 = b^3(-b+2),\quad a_3 = b^4(2b^2-6b+5).
    \label{eq:a2_a3}
\end{equation}
The moments $a_2,a_3$ of Eq.~\eqref{eq:a2_a3} again match the result obtain for the CUE using random-Haar averaging. 
Furthermore, for even higher orders in $1/\lambda$ we check numerically that the moments in the expansion \eqref{eq:Tr_G_expan} match those of the CUE case when $\gamma=2$.
This is consistent with the fact that the random-matrix ensemble \eqref{eq:cayley} used throughout our calculations indeed mimics the CUE when $\gamma=2$.



\bibliographystyle{elsarticle-num} 
\bibliography{meas_ind_trans_draft}






\end{document}